\documentclass[prd,showpacs,preprintnumbers]{revtex4} % Physical Review D

\usepackage{graphicx}% Include figure files
\usepackage{dcolumn}% Align table columns on decimal point
\usepackage{bm}% bold math
\usepackage{latexsym}
\usepackage{amsfonts}
\usepackage{amssymb}
\usepackage{amsmath}
\usepackage{color}
\begin{document}

\preprint{KUNS-2257, DCPT-10/11}

\title{ The Nature of Primordial Fluctuations from Anisotropic Inflation}

\author{Masa-aki Watanabe$^{1)}$}
\author{Sugumi Kanno$^{2)}$}
\author{Jiro Soda$^{1)}$}
\affiliation{1) Department of Physics,  Kyoto University, Kyoto, 606-8501, 
Japan}
\affiliation{2) Centre for Particle Theory, Department of Mathematical 
Sciences, Durham University, Science Laboratories, South Road, Durham, 
DH1 3LE, United Kingdom}

\date{\today}% It is always \today, today,
             %  but any date may be explicitly specified

%===============================================================%
%************************* ABSTRACT ****************************%
%===============================================================%
\begin{abstract}
We study the statistical nature of primordial fluctuations from an anisotropic 
inflation which is realized by a vector field coupled to an inflaton.
We find a suitable gauge, which we call the canonical gauge,
 for anisotropic inflation by generalizing
the flat slicing gauge in conventional isotropic inflation. 
Using the canonical gauge, we reveal the structure of the couplings
between curvature perturbations, vector waves, and gravitational
waves. We identify two sources of anisotropy, i.e. the anisotropy
due to the anisotropic expansion of the universe and that due to
the anisotropic couplings among variables. 
It turns out that the latter effect is dominant. Since the coupling between
the curvature perturbations and vector waves is the strongest one,
the statistical anisotropy in the curvature perturbations is larger than
that in gravitational waves. We find the cross correlation between
the curvature perturbations and gravitational waves
which never occurs in conventional inflation.
We also find the linear polarization of gravitational waves.  
Finally, we discuss cosmological implication of our results. 
\end{abstract}

% PACS, the Physics and Astronomy
                             % Classification Scheme.
%\keywords{Suggested keywords}%Use showkeys class option if keyword
                              %display desired
\maketitle

%===============================================================%
%************************ SECTION I ****************************%
%===============================================================%
\section{Introduction}

The primordial fluctuations from inflation is 
supposed to be statistically isotropic, Gaussian, and scale invariant.
The nature of fluctuations is associated with the nature of
de Sitter spacetime. However,
since the expansion during inflation is not exactly de Sitter, 
the power spectrum is slightly tilted by the order of the slow roll
 parameter~\cite{Komatsu:2010fb}
which characterizes the deviation of the expansion from the exact de Sitter expansion.
The deviation from the Gaussianity is also known to be related to
the slow roll parameter~\cite{Maldacena:2002vr}.
 On the other hand, the statistical isotropy has been
regarded as a robust prediction so far because the cosmic no-hair conjecture
is thought to be robust~\cite{Wald:1983ky}. 

From an observational point of view, there are various indications 
that there exists statistical anisotropy in the cosmic microwave 
background radiation (CMB)~\cite{Eriksen:2003db}. 
Although the statistical significance of these anomalies is still
 under debate, the possibility of the statistical anisotropy
certainly deserves further theoretical investigation~\cite{Gordon:2005ai}.
Recently, breaking the statistical isotropy through the vector fields 
in an inflationary universe is proposed in the paper~\cite{Yokoyama:2008xw}
 and extended in various ways~\cite{Dimopoulos:2009vu,
 Dimastrogiovanni:2010sm,ValenzuelaToledo:2009af}.
 However, if the vector field is relevant to inflation, it may also produce
 anisotropy in an inflationary universe whatever small it is,
 which seems to contradict the cosmic no-hair conjecture.

From the above perspective, it is interesting to ask if it is possible to
have anisotropic inflationary universe~\cite{Ford:1989me,Kaloper:1991rw,
Kawai:1998bn,Barrow:2005qv,Barrow:2009gx,Campanelli:2009tk,Golovnev:2008cf,
Kanno:2008gn}. If possible, it provides a simple mechanism to break the statistical 
isotropy by breaking the isotropy of the spacetime~\cite{Ackerman:2007nb}.
 In the light of no-hair conjecture~\cite{Wald:1983ky}, one may
deny this possibility. In fact, many attempts to construct anisotropic
inflationary models suffer from the instability~\cite{Himmetoglu:2008zp}.   
However, recently, stable anisotropic inflationary
models are found for the first time~\cite{Watanabe:2009ct,Kanno:2009ei}. 
This can be regarded as a counter example to the cosmic no-hair conjecture.
The interesting point is that
the deviation from isotropy is related to the slow roll parameter,
namely, the deviation from the exact deSitter expansion.
Of course, that means the degree of the anisotropy is quite small. 
  From the point of view of precision cosmology, however, 
it is worth exploring theoretical fine structure in an inflationary scenario.

In this paper, we study cosmological perturbations in an anisotropic
 inflationary scenario we have found. 
 The expected phenomenology of the anisotropic inflation is as follows:
\begin{itemize}
\item There should be statistical anisotropy in curvature perturbations.
\item There should be statistical anisotropy in gravitational waves.
\item There should exist the cross correlation 
between curvature perturbations and gravitational waves.
\item There should be linear polarization of gravitational waves.
\end{itemize} 
The first item will be tested by the PLANCK~\cite{Pullen:2007tu}.
The second one may be detected through B-mode polarization
 in the CMB~\cite{Baumann:2008aq}.
The third one will imply T-B correlation in CMB~\cite{Gluscevic:2010vv}.
 The last one could be important for
the future direct measurement of gravitational waves through the
interferometer~\cite{Seto:2001qf}. The purpose of this paper is to
 calculate the above quantities numerically and analytically
 and reveal the physics behind them. 
Since the spacetime is anisotropic, the formalism treating perturbations
is non-standard. Although there are many works treating the cosmological 
perturbations in an anisotropic universe~\cite{Tomita:1985me,Dunsby:1993fg,Noh:1987vk,
Pereira:2007yy,Gumrukcuoglu:2007bx,Himmetoglu:2009mk},
there have been several obstructions in extracting concrete predictions
for CMB. The main obstruction was the lack of the concrete anisotropic cosmological
models. Now, since we have such models, 
we have succeeded in obtaining concrete results by utilizing
 the canonical gauge which is a generalization of the flat slicing
 in the conventional isotropic inflationary scenario.  

Recently, during our slow preparation of this paper, two papers have appeared 
on the archive~\cite{Dulaney:2010sq,Gumrukcuoglu:2010yc}.
The first one \cite{Dulaney:2010sq} studied the primordial perturbations 
in an anisotropic inflationary
universe using a perturbative method.
The second one \cite{Gumrukcuoglu:2010yc}
 investigated the same issue numerically. The conclusion is
quite similar to ours. The main difference is the gauge used in analysis.
Our canonical gauge allows us to reveal the nature of primordial fluctuations
from anisotropic inflation in a transparent way. 
 Interestingly, on the contrary to a naive expectation, all of these works
 including ours imply 
that even if the anisotropy of the universe is very small,  a large
statistical anisotropy in the spectrum of curvature perturbations
could be created.

The organization of this paper is as follows: In section II, we review an
anisotropic inflation which is caused by the inflaton coupled to the vector
field. Here, we will see the anisotropy is determined by the slow roll parameter.
In section III, we choose the canonical gauge which is a generalization
of the flat slicing in the conventional inflation and classify perturbations 
in anisotropic universe based on the 2-dimensional rotation symmetry. 
 Then, we obtain the quadratic action for perturbed quantities.
In section IV, we reduce the action to that for physical variables
from which we can read off the structure of couplings between those variables.
Based on the reduced action, we calculate various statistical quantities numerically
and analytically to reveal
the nature of primordial fluctuations in anisotropic inflation. 
In section V, we discuss cosmological implication of our results.  
The final section is devoted to the conclusion. In the Appendix A,
we provide a detailed derivation of the action for 2-dimensional scalar 
sector perturbations.

%===============================================================%
%************************ SECTION II ****************************%
%===============================================================%
\section{Review of anisotropic inflation}
\label{sc:background}

In this section, we review background solutions proposed 
in \cite{Watanabe:2009ct}, and see how the anisotropic inflation is realized.

We consider the vector field $A_{\mu}$ whose kinetic term is 
coupled to the inflaton field $\phi$. 
We note that this kind of model is quite natural in the context of
the supergravity~\cite{Martin:2007ue}. 
The action is given by
\begin{eqnarray}
S&=&\int d^4x\sqrt{-g}\left[~\frac{1}{2\kappa^2}R
-\frac{1}{2}\left(\partial_\mu\phi\right)\left(\partial^{\mu}\phi\right)
-V(\phi)-\frac{1}{4} f(\phi )^2 F_{\mu\nu}F^{\mu\nu}  
~\right] \ ,
\label{eq:action1}
\end{eqnarray}
where $\kappa ^2$ is the reduced gravitational constant, $g$ is the determinant of the metric, $R$ is the Ricci scalar, $V(\phi)$ is the inflaton potential, $F_{\mu\nu}$ is the field strength of the vector field defined by $F_{\mu\nu}=\partial_{\mu}A_{\nu}-\partial_{\nu}A_{\mu}$, and $f(\phi)$ is a coupling function of the vector field.
We assume that the background spacetime is given by the Bianchi 
type I metric $ds^2 = -dt^2+\sum^3_{i=1}a_i(t)^2dx^{i2}.$
As for the vector field, it can be reduced into "electric"($F_{0i}$) and "magnetic"($F_{ij}$) components, and here we consider only the "electric" component for simplicity.  It is not hard to prove that the direction of the "electric" field does not change in time by solving its evolution equation. Without loosing the generality, one can take  
$x$-axis in the direction of the "electric" field. Using the gauge invariance,
we can express the vector field as $A_{\mu}dx^{\mu} = v(t) dx $. 
Thus, there exists the rotational symmetry in the $y$-$z$ plane. 
 Given this configuration, it is convenient to parameterize the metric as follows:
\begin{equation}
ds^2 = -{\cal N}(t)^2dt^2 + e^{2\alpha(t)} \left[ e^{-4\sigma (t)}dx^2
+e^{2\sigma (t)}(e^{2\sqrt{3}\sigma_{-}(t)}dy^2
+e^{-2\sqrt{3}\sigma_{-}(t)}dz^2) \right] \ ,
\end{equation} 
where $e^\alpha$, $\sigma$ and $\sigma_{-}$ are an isotropic scale factor  
 and spatial shears, respectively.
We also define the averaged Hubble parameter as $H \equiv \dot{\alpha}$.
 Here, the lapse function ${\cal N}$ is introduced to obtain the Hamiltonian constraint. 
With the above ansatz, the action becomes
\begin{equation}
S=\int d^4x \frac{1}{\cal N}e^{3\alpha} \left[ \frac{3}{\kappa ^2} (-\dot{\alpha}^2+\dot{\sigma}^2+\dot{\sigma}_{-}^2)+\frac{1}{2}\dot{\phi}^{2}-{\cal N}^2 V(\phi)+\frac{1}{2}f(\phi )^2\dot{v}^2 e^{-2\alpha (t) +4\sigma(t) } \right],
\end{equation}
where an overdot denotes the derivative with respect to the physical time $t$.
First, its variation with respect to $\sigma_{-}$ yields
\begin{equation}
\ddot{\sigma}_{-}=-3\dot{\alpha}\dot{\sigma}_{-} \ .
\end{equation}
This gives $\dot{\sigma}_{-}\propto e^{-3\alpha}$, hence,
 the anisotropy in the $y$-$z$ plane rapidly decays as the universe expands. 
 Hereafter, for simplicity, we assume $\sigma_{-}=0$ and set the metric to be 
\begin{eqnarray}
 ds^2 = -dt^2 + e^{2\alpha(t)} \left[ e^{-4\sigma (t)}dx^2
 +e^{2\sigma (t)}(dy^2+dz^2) \right] \ .
\end{eqnarray}
Next, the equation of motion for $v$ is easily solved as
\begin{equation}
\dot{v} =f(\phi)^{-2}  e^{-\alpha -4\sigma}  p_A  \ ,
\label{eq:Ax}
\end{equation}
where $p_A$ is a constant. Taking the variation of the action with respect to 
${\cal N}, \alpha, \sigma$ and $\phi$ and substituting the solution (\ref{eq:Ax}), 
we obtain the following basic equations:
\begin{eqnarray}
\dot{\alpha}^2 &=& \dot{\sigma}^2+\frac{\kappa ^2}{3} \left[ \frac{1}{2}\dot{\phi}^2+V(\phi )+\frac{p_A^2}{2}f(\phi )^{-2}e^{-4\alpha -4\sigma} \right] \ , \label{eq:hamiltonian}\\
\ddot{\alpha} &=& -3\dot{\alpha}^2+\kappa ^2 V(\phi ) +\frac{\kappa ^2 p_A^2}{6}f(\phi )^{-2}e^{-4\alpha-4\sigma} \ , \label{eq:alpha}\\
\ddot{\sigma} &=&  -3\dot{\alpha}\dot{\sigma}+\frac{\kappa ^2 p_A^2}{3}f(\phi)^{-2}e^{-4\alpha -4\sigma} \ , \label{eq:sigma}\\
\ddot{\phi} &=& -3\dot{\alpha}\dot{\phi}-V_{\phi}+p_A^2f(\phi)^{-3} f_{\phi}e^{-4\alpha -4\sigma} \label{eq:inflaton} \ ,
\end{eqnarray}
where the subscript in $V_\phi$ denotes derivative with respect to $\phi$.
 Let us check whether inflation occurs in this model. Using Eqs. (\ref{eq:hamiltonian}) and (\ref{eq:alpha}), the equation for acceleration of the cosmic expansion is given by
\begin{equation}
\frac{(e^{\alpha})^{\cdot \cdot}}{e^{\alpha}} = \ddot{\alpha}+\dot{\alpha}^2 = -2\dot{\sigma}^2-\frac{\kappa ^2}{3} \dot{\phi}^2 + \frac{\kappa ^2}{3} \left[ V - \frac{p_A^2}{2}f^{-2}e^{-4\alpha -4\sigma} \right] \ .
\end{equation}
We see that the potential energy of the inflaton needs to be dominant and the energy density of the vector field $\rho _v \equiv p_A^2 f(\phi)^{-2}e^{-4\alpha -4\sigma}/2$ 
and the shear $\Sigma \equiv \dot{\sigma}$ should be subdominant for inflation to occur. Next we want to see if the anisotropy is produced during inflation. Here we look at the ratio of the shear to the expansion rate $\Sigma/H$ to
 characterize the anisotropy of the inflationary universe. 
 Then, Eq.(\ref{eq:sigma}) reads 
\begin{eqnarray} 
 \dot{\Sigma} = -3H\Sigma+\frac{2\kappa ^2}{3}\rho _v \ .
\end{eqnarray}
 If the anisotropy converges to a value, i.e. $\dot{\Sigma}$ becomes negligible,
 the terminal value should be given by 
\begin{equation}
\frac{\Sigma}{H} = \frac{2}{3}{\cal R} \ ,
\end{equation}
where  we used the slow roll equation
 $H^2 = \kappa ^2 V(\phi) /3$ derived from Eq.(\ref{eq:hamiltonian}) and
 defined the energy ratio ${\cal R} \equiv \rho _v / V(\phi) $.

In order to realize the above situation, $\rho_v$ must be almost constant.
Assuming that the vector field is subdominant in the evolution equation of 
the inflaton field Eq.(\ref{eq:inflaton}) and the conventional single field slow-roll inflation is realized, one can show the coupling function $f(\phi)$ should be
proportional to $e^{-2\alpha}$ to keep $\rho_v$ almost constant. 
 In the slow roll phase, $e$-folding number $\alpha$ is related to the inflaton 
 field $\phi$ as $d\alpha = - \kappa ^2 V(\phi) d\phi / V_\phi$ as usual.  
Then,  the functional form of $f(\phi)$ is determined as
\begin{equation}
f(\phi) = e^{-2\alpha} = e^{2\kappa ^2 \int \frac{V}{V_\phi} d\phi} \ . 
\label{eq:function}
\end{equation}
For the polynomial potential $V\propto \phi ^n$, for example, we have $f=e^{\frac{\kappa ^2 \phi ^2}{n}}$. In this scenario, the anisotropy is restricted from the condition that the vector field is negligible in Eq.(\ref{eq:inflaton}), that is, $|p_A^2f(\phi)^{-3}f_{\phi}e^{-4\alpha -4\sigma}| \ll |V_\phi|.$ 
Substituting Eq.(\ref{eq:function}) into this, we obtain 
${\cal R} \ll \epsilon_{ V}/2,$
where $\epsilon_{V}$ is the slow-roll parameter defined by 
$\epsilon_{V} \equiv 1/2\kappa ^2 \left( V_\phi /V \right) ^2$. 
Thus, the anisotropy is constrained by $\Sigma/H \ll {\cal O}(\epsilon_V)$.

The above case is, in a sense, a critical one, 
and next we want to consider beyond it. 
For simplicity, we parameterize $f(\phi)$ by 
\begin{equation}
f(\phi) = e ^{2c\kappa ^2 \int \frac{V}{V_\phi}d\phi},
\end{equation}  
where $c$ is a constant parameter. Now, we look at what happens when $c>1$. 
The basic equations become
\begin{eqnarray}
\frac{(e^{\alpha})^{\cdot \cdot}}{e^{\alpha}} 
&=& \ddot{\alpha}+\dot{\alpha}^2 
= -2\dot{\sigma}^2-\frac{\kappa ^2}{3} \dot{\phi}^2 
+ \frac{\kappa ^2}{3} V(\phi) \left[ 1-{\cal R} \right] \ , \\
\ddot{\phi} &=& -3\dot{\alpha}\dot{\phi}-V_\phi \left[ 1-\frac{2c}{\epsilon_V}
 {\cal R} \right] \ . \label{eq:inflaton2}
\end{eqnarray}
 In this case, if the vector field is initially small 
 ${\cal R} \ll \epsilon_V /2c $, then the conventional single field slow-roll inflation is realized. During this stage $f\propto e^{-2c\alpha}$
  and the vector field grows as $\rho_v \propto e^{4(c-1)\alpha}$. Therefore, the vector
  field eventually becomes relevant to the inflaton dynamics Eq.(\ref{eq:inflaton2}). 
  Nevertheless, the accelerating expansion of the universe will continue. 
 The point is that  ${\cal R}$ cannot exceed $\epsilon_{V}/2c$.
 In fact, if ${\cal R}$ exceeds $\epsilon_{V}/2c$, the inflaton field $\phi$
 does not roll down, which makes 
 $\rho _v = p_A^2 f(\phi)^{-2}e^{-4\alpha-4\sigma} /2$
 decrease. Hence,  $\rho_v \ll V(\phi)$ always holds. 
 In this sense, there exists an attractor where
 the inflation continues even when the vector field affects the inflaton evolution. 
The inflaton dynamics is determined by solving the slow-roll equation:
\begin{equation}
 -3\dot{\alpha}\dot{\phi}-V_{\phi}+p_A^2f^{-3}f_{\phi}e^{-4\alpha -4\sigma} =0. 
 \label{eq:inflaton3}
 \end{equation}
Assuming $\dot{\alpha}^2= \kappa ^2 V(\phi)/3 $, this yields
\begin{equation}
\frac{d \phi}{d \alpha}=\frac{\dot{\phi}}{\dot{\alpha}}=-\frac{V_\phi}{\kappa ^2 V} + 2c \frac{p_A^2}{V_\phi}e^{-4\alpha -4\sigma -4c \kappa^2 \int\frac{V}{V_\phi}d\phi}
\label{eq:inflaton4}
\end{equation}
This can be integrated by neglecting the evolutions of $V,V_\phi,\sigma$ as
\begin{eqnarray}
  e^{4\alpha +4\sigma +4c \kappa^2 \int\frac{V}{V_\phi}d\phi}
  = \frac{2c^2 p_A^2}{c-1} \frac{\kappa^2 V}{V_{\phi}^2} 
  \left[ 1+ {\rm const.} e^{-4(c-1)\alpha +4\sigma} \right]  \ .
\end{eqnarray}
Substituting this into the slow-roll equation Eq.(\ref{eq:inflaton4}), we obtain
\begin{eqnarray}
\frac{d\phi}{d\alpha} &=& - \frac{V_{\phi}}{\kappa ^2 V}+\frac{c-1}{c}\frac{V_{\phi}}{\kappa ^2V} \left[ 1+ {\rm const.}  e^{-4(c-1)\alpha+4\sigma}   \right] ^{-1} \ .
\end{eqnarray}
This clearly shows a transition from the conventional single field slow-roll inflationary phase, where $d\phi /d\alpha =- V_{\phi} / \kappa ^2 V$ holds, to what we refer to as the second inflationary phase, where the vector field is relevant to the inflaton dynamics and the inflaton gets $1/c$ times slower as 
$d\phi / d\alpha = - V_{\phi} /c \kappa^2 V $.
 In the second inflationary phase, the energy density of the vector field
 becomes
\begin{equation}
\rho _v =\frac{p_A^2}{2} e^{-4\alpha-4\sigma-4c \kappa^2 \int\frac{V}{V_\phi}d\phi}
 = \frac{1}{2} \frac{c-1}{c^2} \epsilon_{V} V(\phi) \ ,
\end{equation}
which yields the anisotropy of $\Sigma/H= 2{\cal R}/3=(c-1)\epsilon_{V}/3c^2$.
 From Eqs.(\ref{eq:hamiltonian}) and (\ref{eq:alpha}), 
 the slow-roll parameter defined in terms of the scale factor becomes
\begin{equation}
\epsilon _{H}\equiv -\frac{\ddot{\alpha}}{\dot{\alpha}^2}=-\frac{1}{\dot{\alpha}^2} \left(-\frac{1}{2} \kappa ^2 \dot{\phi}^2-\frac{2}{3}\kappa ^2 \rho_v \right) = \frac{1}{c} \epsilon_{V} \ ,
\end{equation}
where we neglected the anisotropy and used relations 
$\dot{\phi}/\dot{\alpha} = d\phi /d\alpha = - V_{\phi}/ c \kappa^2 V $
and $\dot{\alpha}^2= \kappa ^2 V(\phi)/3 $.
Thus we have a remarkable result
\begin{equation}
\frac{\Sigma}{H} = \frac{1}{3}\frac{c-1}{c}\epsilon_{H}.
\end{equation}
In the next section, we will make a perturbative analysis during this second inflationary phase.

%===============================================================%
%************************ SECTION III ***************************%
%===============================================================%
\section{Canonical Gauge in Anisotropic Inflation}

Since the background is expanding anisotropically, we cannot use
the standard cosmological perturbation theory. 
In this section, we classify perturbations under the 
2-dimensional rotational symmetry and obtain the quadratic actions
for 2-dimensional scalar and vector sectors. 
In order to grasp the meaning of variables, we start with the isotropic case
and make a gauge transformation
from the flat slicing gauge to the appropriate gauge for 2-dimensional
classification. Then, the resultant gauge can be promoted to the anisotropic 
spacetime. The gauge we have chosen makes the analysis and the interpretation 
easier. Once the gauge is fixed, the quadratic action can be calculated.

\subsection{Gauge Fixing and Classification of perturbations}

First, we will start with the spatially homogeneous and isotropic universe.
For simplicity, we consider flat space. 
\begin{eqnarray}
ds^2 = a^2 (\eta) \left[ -d\eta^2 + \delta_{ij} dx^i dx^j \right] \ ,
\end{eqnarray}
where we took a conformal time $\eta$. 
In that case, we can use 3-dimensional rotational symmetry to classify
the perturbed metric. 
When we want to have diagonal quadratic action, we take the following gauge
\begin{eqnarray}
ds^2 = a^2 \left[ -(1+2A)d\eta^2 +2 (B_{,i}+V_i ) d\eta dx^i
+  (\delta_{ij} + h_{ij} ) dx^i dx^j \right]
\end{eqnarray}
where we imposed $V_{i,i}=0$ and $ h_{ij,j}=h_{ii}=0$.
If we ignore vector and tensor perturbations $V_i , h_{ij}$,
the above gauge is called the flat slicing gauge.
Now, let us move on to the Fourier space. Since there exists 3-dimensional
rotation symmetry, we can take a wave number vector to be
${\bf k} = (k,0,0)$. Then, the perturbed metric has the following components:
\begin{equation}
\delta g_{\mu \nu} =\left(
\begin{array}{ccccc}
& -2 a^2 A~ &  ~a^2 B_{,x}~ & ~a^2 V_2~ & ~a^2 V_3~
\vspace{1mm}\\\vspace{1mm}
& \ast  & 0 &    0            &   0              
\\\vspace{1mm}
& \ast & \ast & a^2 h_{+}      &  -a^2 h_{\times} 
\\
& \ast     & \ast & \ast &  -a^2 h_{+}
\end{array}
\right) \ . \hspace{5mm} * {\rm ~is ~symmetric ~part}.
\end{equation}
Here, we utilized the special choice ${\bf k} = (k,0,0)$ to solve
the constraints $V_{i,i}=0$ and $ h_{ij,j}=h_{ii}=0$. With the same 
reason, only $B_{,x}$ remains. We defined $h_{yz}=-h_{\times}, 
h_{yy}=-h_{zz}=h_{+}$.
Now, we will pretend that we have only 2-dimensional rotation symmetry 
in $y$-$z$ plane. In that case, at best,
we can take ${\bf k}=(k_x, k_y ,0)$. Hence, 
we make a rotation in the $x-y$-plane so that the wave number becomes
${\bf k}=(k_x, k_y ,0)$. 
\begin{eqnarray}
  \left(
\begin{array}{cc}
& k_x \\
& k_y \\
& 0  
\end{array}
\right)
= \frac{1}{k}\left(
\begin{array}{cccc}
& k_x~ & ~-k_y~ & ~0~ \\\vspace{1mm}
& k_y & k_x & 0 \\
& 0 & 0 & 0
\end{array}
\right)
 \left(
\begin{array}{cc}
& k \\
& 0 \\
& 0   
\end{array}
\right)  
\ ,
\end{eqnarray}
where we have a relation $k^2 = k_x^2 + k_y^2$.
Under this rotation, the perturbed metric becomes
\begin{equation}
\delta g_{\mu \nu} =\left(
\begin{array}{ccccc}
& -2 a^2 A~ & ~\frac{k_x}{k} a^2 B_{,x} -\frac{k_y}{k} a^2 V_2~ 
& ~\frac{k_y}{k} a^2 B_{,x} +\frac{k_x}{k} a^2 V_2~          
& a^2 V_3~ 
\vspace{2mm}\\\vspace{2mm}
& \ast &  a^2 \frac{k_y^2}{k^2}h_{+} 
& -a^2 \frac{k_x k_y}{k^2}  h_{+} & a^2 \frac{k_y}{k} h_{\times}  
\\\vspace{2mm}
& \ast & \ast    
& a^2 \frac{k_x^2}{k^2} h_{+} &  -a^2 \frac{k_x}{k} h_{\times} 
\\
& \ast & \ast & \ast &  -a^2 h_{+}
\end{array}
\right) \ .
\end{equation}
To simplify the perturbations, we can make use of gauge transformation
\begin{eqnarray}
\delta g_{\mu\nu} \rightarrow 
\delta g_{\mu\nu} + \xi_{\mu ; \nu} +\xi_{\nu;\mu} \ ,
\end{eqnarray}
where the semicolon denotes the covariant derivative with respect to the background
metric. Taking the parameter 
$$
\xi^0 =0 \ , \quad \xi^x = \frac{k_x}{2i k^2 }h_{+} \ , \quad
\xi^y = \frac{k_y}{2i k^2} h_{+} \ , \quad
\xi^z = \frac{k_x}{ik_y k } h_{\times} \ ,
$$
we obtain
\begin{equation}
\delta g_{\mu \nu} =\left(
\begin{array}{ccccc}
& -2 a^2 A~ 
& ~\frac{k_x}{k}a^2B_{,x}-\frac{k_y}{k}a^2V_2+\frac{k_x}{2i k^2 }a^2h'_{+}~ 
& ~\frac{k_y}{k}a^2 B_{,x}+\frac{k_x}{k}a^2V_2+\frac{k_y}{2i k^2}a^2h'_{+}~ 
& ~a^2 V_3+\frac{k_x}{ik_y k}a^2h'_{\times}~  
\vspace{2mm}\\\vspace{2mm}
& \ast &  a^2 h_{+}  
&  0  &   a^2 \frac{k}{k_y} h_{\times}  
\\\vspace{2mm}
& \ast &  \ast    
&  a^2  h_{+}   &  0 
\\
& \ast & \ast 
& \ast &  -a^2 h_{+}
\end{array}
\right) \ .
\label{gauge-A}
\end{equation}
It should be noted that we did not change slicing but
 performed only the spatial coordinate transformation. Therefore, 
 we are still working in the flat slicing where the 3-dimensional scalar curvature
 vanishes. 
 
In our anisotropic inflation models, the available symmetry is actually small. 
The background metric is given by
\begin{equation}
ds^2_b =a(\eta) ^2(-d\eta ^2+dx^2)+b(\eta) ^2(dy^2+dz^2),
\end{equation}
that is, $ a=e^{\alpha -2\sigma}, b=e^{\alpha +\sigma}, d\eta=dt/a$.
Notice that the conformal time in anisotropic inflation is the conformal
time in 2-dimensional part $(\eta , x)$. 
Even in this anisotropic spacetime,
as we have done in (\ref{gauge-A}), one can take the following gauge
\begin{equation}
\delta g_{\mu \nu} =\left(
\begin{array}{ccccc}
& \delta g_{00}~ & ~\delta g_{0x}~ & ~\delta g_{0y}~ & ~\delta g_{0z}~
\vspace{1mm}\\\vspace{1mm}
& *  &  \delta g_{xx}  &  0 &  \delta g_{xz} 
\\\vspace{1mm}
&  * &  * &  \delta g_{yy}  &  0 
\\\vspace{1mm}
&  *  &  *   & * &  \delta g_{zz}
\end{array}
\right) \ ,
\end{equation}
where we can impose further conditions so that the perturbed metric goes back
to (\ref{gauge-A}) in the isotropic limit. 

One can classify the perturbed metric using the rotational symmetry in $y-z$-plane.
In 2-dimensional flat space, an arbitrary vector $m^a $ where $a=y,z$ can be 
decomposed into the scalar part $m^a_{,a}\neq 0$ and the vector part $m^a_{,a} =0$. 
Since there exists no tensor part in 2-dimensions, 2-dimensional tensor can be
constructed from the 2-dimensional vector. 
Because of the symmetry, the scalar and vector parts are not mixed in the equations.
Thus, the metric perturbations
can be classified to the scalar sector and the vector sector. 
Thanks to the symmetry in the $y-z$ plane,  without loss of generality, 
 we can take the wave number vector to be ${\bf k} = (k_x , k_y ,0)$. 
Hence, the vector sector in 2-dimensional classification can be identified as
$\delta g_{0z} , \delta g_{xz}$ in the above perturbed metric.
The remaining components $\delta g_{00} , \delta g_{0x} , \delta g_{0y} ,
\delta g_{xx}, \delta g_{yy}, \delta g_{zz} $ belong to the scalar sector.

\subsection{2d vector sector}

 Thus, the perturbations that belong to 2d vector perturbations, 
 can be written as
\begin{equation}
\delta g_{\mu \nu}^{\rm vector} =\left(
\begin{array}{ccccc}
& 0~ & ~0~ & ~0~ & ~b^2 \beta_3~ \vspace{1mm}\\\vspace{1mm}
& *  &  0  &  0 & b^2 \Gamma \\\vspace{1mm}
&  * &  * &  0  &  0 \\
&  *  &  *   & * &  0
\end{array}
\right) \ ,
\end{equation}
where we have incorporated the anisotropy while keeping the spatial 
scalar curvature to be zero. 
As to the vector field, we can take 
\begin{eqnarray}
\delta A_{\mu}^{\rm vector} = \left(0 \ , 0 \ , 0 \ ,  D \right) \ .
\end{eqnarray}
Note that we have no residual gauge transformation and, in particular,
 $D$ is a gauge invariant under abelian gauge transformations. 
 And, as we have seen in (\ref{gauge-A}), $\Gamma$ corresponds to
  the cross-mode polarization of gravitational waves in the isotropic limit $a=b$.

Using this gauge, we can calculate the quadratic action as
\begin{eqnarray}
S^{{\rm vector }}
&=&
\int d\eta d^3 x \left[~ 
\frac{b^4}{4a^2} \beta^2_{3,x} + \frac{b^2}{4} \beta^2_{3,y} 
- \frac{b^4}{2a^2} \Gamma' \beta_{3,x} 
+ \frac{f^2 v' b^2}{a^2} \beta_3 D_{,x}     \right. \nonumber \\
&&  \left. \qquad\qquad
-\frac{b^2}{4} \Gamma^2_{,y} + \frac{b^4}{4a^2} \Gamma^{\prime 2} 
-\frac{f^2a^2}{2b^2} D_{,y}^2
-\frac{1}{2}f^2 D_{,x}^{2}+\frac{f^2}{2} D^{\prime 2}
-\frac{f^2 v' b^2}{a^2} D' \Gamma ~\right] \ .
\end{eqnarray}
Since the perturbed shift function $\beta_3$ does not have a time derivative,
  it is not dynamical. 
There are two physical degrees of freedom $\Gamma$ and $D$ in this
2-dimensional  vector sector. 

\subsection{2d scalar sector}

For the 2-dimensional scalar sector, we define the metric perturbations
\begin{equation}
\delta g_{\mu \nu} =\left(
\begin{array}{ccccc}
& -2 a^2 \Phi~ & ~a \beta_1~  & ~a \beta_2~ & ~0~ \vspace{2mm}\\\vspace{1mm}
& *  &  2 a^2 G  &  0 & 0 \\\vspace{1mm}
&  * &  * &  2 b^2 G  &  0 \\
&  *  &  *   & * & -2 b^2 G
\end{array}
\right) \ ,
\end{equation}
where we have kept the spatial scalar curvature vanishing. 
The scalar perturbation will be represented by $\delta \phi$. 
The variable $G$ and $\delta \phi$ are the gauge invariant variables 
that correspond to the plus mode of gravitational waves 
and the scalar perturbations, respectively, in the isotropized  limit $a=b$. 
And, we set the perturbed vector to be
\begin{eqnarray}
\delta A^{\rm scalar}_{\mu} = \left( \delta A_0 \ , 0 \ ,   J \ , 0 \right)  \ ,
\end{eqnarray} 
where we have fixed the abelian gauge by putting the longitudinal component 
to be zero.
From these ansatz, we can calculate the quadratic action as
\begin{eqnarray}
 S^{{\rm scalar}}
  &=& \int d^3 x d\eta 
 \left[~ 
 \frac{b^2}{2a^2} f^2 \delta A_{0,x}^2 + \frac{f^2}{2} \delta A_{0,y}^2
 +\frac{b^2}{a^2} f^2 v' \left(G+\Phi \right) \delta A_{0,x} -f^2 J' \delta A_{0,y}
  -2 \frac{b^2}{a^2} ff_\phi v'\delta \phi \delta A_{0,x} 
   \nonumber \right. \\ 
&& +\frac{1}{4} \beta_{1,y}^2
 -\frac{1}{2} \beta_{2,x} \beta_{1,y} + 2\frac{bb'}{a} \Phi_{,x} \beta_1 
 - \frac{b^2}{a} \phi' \delta\phi_{,x} \beta_1 + \frac{1}{4} \beta_{2,x}^2 
  + a \left( \frac{a'}{a}+\frac{b'}{b} \right) \beta_2 \Phi_{,y} \nonumber \\
&& -a \left(\frac{a'}{a} -\frac{b'}{b} \right) \beta_2 G_{,y} 
 +  \frac{f^2}{a} v' \beta_2 J_{,x} -a \phi' \beta_2 \delta\phi_{,y} 
  + \frac{1}{2} f^2 J^{\prime 2} -\frac{1}{2} f^2 J_{,x}^2 
   +b^2 G^{\prime 2} -a^2 G_{,y}^2 -b^2 G_{,x}^2 
   \nonumber\\
&&  +\frac{1}{2}b^2 \delta\phi^{\prime 2}
 -\frac{a^2}{2} \delta\phi_{,y}^2 - \frac{b^2}{2} \delta\phi_{,x}^2
 -\frac{1}{2} a^2 b^2 V_{\phi\phi} \delta\phi^2 
 + \frac{b^2}{2a^2} \left( f_\phi^2 +ff_{\phi\phi} \right) 
 v^{\prime 2} \delta\phi^2          
 - a^2 b^2 V  \Phi^2  \nonumber \\
&&  \left.
 +  \frac{b^2}{2 a^2} f^2 v^{\prime 2}  G^2 
 - 2a^2 b^2 V \Phi G    -2bb' \Phi' G  
 - \left( \frac{b^2}{a^2} ff_\phi v^{\prime 2} + a^2 b^2 V_\phi \right)
  \delta\phi \left( G+\Phi \right) +b^2 \phi' \delta\phi' \left( G-\Phi \right)
  ~\right]
\end{eqnarray}
Here, $S^{{\rm scalar}}$ consists of $\Phi, \beta_1, \beta_2, G, \delta A_0 , \delta \phi$
 and $J $. Among them, $\Phi , \beta_1, \beta_2$ and $\delta A_0 $ 
 are non-dynamical and can be eliminated.

%%%%%%%%%%%%%%%%%%%%%%%%%%%%%%%%%%%%%%%%%%%%%%%%%%%%%%%%%%%%%%%%%%
%%%%%%%%%%%%%%%%%%%%%%% Section IV %%%%%%%%%%%%%%%%%%%%%%%%%%%%%%%
%%%%%%%%%%%%%%%%%%%%%%%%%%%%%%%%%%%%%%%%%%%%%%%%%%%%%%%%%%%%%%%%%%
\section{The nature of Primordial Fluctuations}

Now, we are in a position to calculate the statistical properties
of primordial fluctuations from anisotropic inflation. 
For this purpose, we need to reduce the action to the one for physical variables.
Then, we can quantize the system and specify the vacuum state.
We analyze the vector sector and the scalar sector separately. 

%%%%%%%%%%%%%%%%%%%%%%%%%%%%%%%%%%%%%%%%%%%%%%%%%%%%%%%%%%%%%%%%%%
%%%%%%%%%%%%%%%%%%%%%%%%%%%%%%%%%%%%%%%%%%%%%%%%%%%%%%%%%%%%%%%%%%
%%%%%%%%%%%%%%%%%%%%%%%%%%%%%%%%%%%%%%%%%%%%%%%%%%%%%%%%%%%%%%%%%%

\subsection{Action in slow roll approximation}

First, let us eliminate non-dynamical variables from the action for
the 2-dimensional vector sector.
In Fourier space, the action for 2-dimensional vector sector becomes
\begin{eqnarray}
S^{{\rm vector}}&=&
\int d\eta d^3k \left[~ 
\frac{1}{4} \left( \frac{b^4}{a^2} k_x^2 |\beta_3|^2 +b^2 k_y^2 |\beta_3|^2 
\right) 
+ik_x\frac{b^4}{2a^2}\beta_3^{*} \Gamma^{'} 
-ik_x\frac{f^2v^{'}b^2}{a^2}\beta_3  D^{*} \right.\nonumber\\
&&  \left. \qquad\qquad
+\frac{b^4}{4a^2}|\Gamma'|^2 - \frac{b^2}{4} k_y^2 |\Gamma|^2 
 +\frac{f^2}{2}|D^{'}|^2 -\frac{f^2a^2}{2b^2} k_y^2 |D|^2
-\frac{f^2}{2} k_x^2 |D|^{2}
-\frac{f^2v^{'}b^2}{a^2}D^{'}\Gamma^{*} ~\right] \ ,
\end{eqnarray}
where we omitted the Fourier indices ${\bf k}$ for simplicity. 
After making the above action real manifestly and 
completing the square of $\beta_3$, we obtain
\begin{eqnarray}
L^{\rm vector}&=&
\frac{b^4}{4a^2}k^2 \left|~ \beta_3
+i\frac{k_x}{k^2}\Gamma^\prime
+2i\frac{f^2}{b^2}\frac{k_x}{k^2}v^\prime D~\right| ^2 
 +\frac{b^2}{4}\frac{k_y^2}{k^2}|\Gamma^{'}|^2-\frac{b^2}{4}k_y^2|\Gamma|^2 \nonumber\\
&\ & +\frac{f^2}{2}|D^{'}|^2-\frac{f^2}{2}k^2|D|^2
-\frac{f^4v^{'2}}{a^2}\frac{k_x^2}{k^2}|D|^2 
 -\frac{f^2v^{\prime}}{2}\frac{k_y^2}{k^2}\left(\Gamma^{'}D^{*}
 +\Gamma^{*{'}}D\right)  \ ,
\end{eqnarray}
where $k$ is given by $k(\eta) \equiv \sqrt{k_x^2+ a^2 k_y^2/b^2}$
which becomes constant in the isotropic limit $a=b$.
The first squared term vanishes after substituting the equation of motion for 
$\beta_3$.
Now, we define canonically normalized variables as
\begin{eqnarray}
\bar{\Gamma}  \equiv  \frac{b |k_y|}{\sqrt{2}k}\Gamma, \qquad
\bar{D}  \equiv  fD.
\end{eqnarray}
Then, using the canonical variables and integrating by parts, 
we obtain the reduced action for physical variables
\begin{eqnarray}
S^{\rm vector} &=& \int d\eta d^3 k \left[ 
\frac{1}{2} |\bar{\Gamma}^{'}|^2
+\frac{1}{2}\left( \frac{(b/k)^{''}}{(b/k)}-k^2 \right) |\bar{\Gamma}|^2 
 +\frac{1}{2}|\bar{D}^{'}|^2
 +\frac{1}{2}\left( \frac{f^{''}}{f}-k^2
 -2\frac{f^2v^{'2}}{a^2}\frac{k_x^2}{k^2} 
                         \right) |\bar{D}|^2  \right. \nonumber\\
 &&\qquad \qquad \qquad \left. 
 + \frac{1}{\sqrt{2}} \frac{fv^{'}}{a}\frac{a}{b}\frac{k_y}{k} 
 \left\{ \bar{\Gamma}^{'}\bar{D}^{*} +  \bar{\Gamma}^{*'}\bar{D}
 +\frac{(k/b)^{'}}{(k/b)}\left( \bar{\Gamma}\bar{D}^{*} + \bar{\Gamma}^{*} \bar{D} \right)
                              \right\} \right]  \ .
\label{vec-action}
\end{eqnarray}
In the isotropic limit $a=b$,
 $\bar{\Gamma} $ and $\bar{D}$ represent the cross-mode of gravitational waves
and vector waves, respectively. The second line in the action (\ref{vec-action})
describes how both waves are interacting to each other. 

Next, we use the slow roll approximation to simplify the action.
To obtain the homogeneous background metric, 
we integrate the following equations
\begin{eqnarray}
-\frac{\dot{H}}{H^2} = \epsilon _H, \qquad
\frac{\Sigma}{H}  = \frac{1}{3} I \epsilon _H \ ,
\end{eqnarray}
by assuming $\epsilon_H^{'}/\epsilon_H \ll a^{'} /a$.
The resultant expressions are 
\begin{eqnarray}
a = (-\eta )^{-1-\epsilon _H }, \qquad
b = (-\eta )^{-1-\epsilon _H - I \epsilon_H } \ .
\end{eqnarray}
In this approximation, the universe shows anisotropic power law inflation.
We should recall, in the second inflationary phase, the variable $I$ is given by
\begin{equation}
I = \frac{c-1}{c} \ .
\end{equation}
Note that the range $(1,\infty )$ for $c$ corresponds to $(0,1)$ for $I$.
The above approximation gives useful formula for the subsequent calculations
\begin{eqnarray}
\frac{a^{'2}}{a^2} 
&=& (-\eta)^{-2}\left[~1+2\epsilon _H~\right] \ , \quad
\frac{a^{''}}{a}=(-\eta)^{-2}\left[~2+3\epsilon _H~\right] \ ,\quad
\frac{a^\prime}{a}\frac{b^\prime}{b}
=(-\eta)^{-2}\left[~1+2\epsilon _H+I\epsilon _H~\right] \ ,
\nonumber\\
\frac{b^{'2}}{b^2} 
&=& (-\eta)^{-2} \left[~1+2\epsilon _H + 2 I\epsilon _H~\right] \ , \quad
\frac{b^{''}}{b} = (-\eta )^{-2} \left[~2+3\epsilon _H + 3 I \epsilon _H~\right] \ , 
\nonumber\\
\frac{k ^{'}}{k} &=& (-\eta )^{-1} \left[~-I\epsilon _H~\right] 
\frac{k_y^2}{k^2}\frac{a^2}{b^2} \ , \quad
\frac{k ^{''}}{k} = (-\eta )^{-2} \left[~-I\epsilon _H~\right] 
\frac{k_y^2}{k^2}\frac{a^2}{b^2} \ . 
\label{parts}
\end{eqnarray}
From the background equations Eqs.(\ref{eq:hamiltonian})-(\ref{eq:inflaton}), 
it is easy to obtain
\begin{eqnarray}
\frac{f^2v^{'2}}{a^2} = \frac{b^{'2}}{b^2} +\frac{b^{''}}{b}+\frac{a^{'2}}{a^2}
-\frac{a^{''}}{a}-2\frac{a^{'}b^{'}}{ab} \ .
\end{eqnarray}
Using the formula (\ref{parts}), we obtain
\begin{eqnarray}
\frac{f^2 v^{'2}}{a^2} &=& 3(-\eta )^{-2} I \epsilon _H \ . 
\end{eqnarray}
From Eq.~(\ref{eq:Ax}), the background equation for the vector can be found as
\begin{eqnarray}
\left[ \frac{f^2v^{'}b^2}{a^2} \right] ^{'} = 0  \ .
\end{eqnarray}
From this equation, it is easy to deduce the relation
\begin{eqnarray}
\frac{f^{'}}{f} = (-\eta )^{-1}\left[ -2 -3\epsilon _H + \eta _H 
-2 I \epsilon _H \right] \ ,
\end{eqnarray}
where $\eta _H$ is a slow-roll parameter defined by
\begin{equation}
\frac{\epsilon _H ^{'}}{\epsilon _H} 
= 2 \frac{(e^{\alpha})^{'}}{e^{\alpha}} \left( 2 \epsilon _H - \eta _H \right) 
= 2 (2 \epsilon _H -\eta _H) (-\eta )^{-1} \ .
\end{equation}
Of course, $\eta_H$ is not related to the conformal time $\eta$. 
Furthermore, we obtain
\begin{eqnarray}
\frac{f^{''}}{f} &=& 
(-\eta )^{-2} \left[ 2+9\epsilon _H -3 \eta _H + 6 I\epsilon _H \right] \ .
\end{eqnarray}
Substituting these results into the action, we obtain the action
in the slow roll approximation
\begin{eqnarray}
S^{\rm vector} &=& \int d\eta d^3 k \left[
 \frac{1}{2} | \bar{\Gamma}^{'} | ^2 
+\frac{1}{2} \left[ -k^2+ (-\eta )^{-2} \left\{ 2+3\epsilon _H+3I \epsilon _H 
+ 3 I\epsilon _H \sin^2 \theta \right\} \right] |\bar{\Gamma}| ^2
\right. \nonumber \\
 && \qquad \qquad + \frac{1}{2}|\bar{D}^{'}|^2 
 +\frac{1}{2} \left[ -k^2+ (-\eta)^{-2} \left\{ 2+9\epsilon _H -3\eta _H 
  +6 I\epsilon _H \sin^2 \theta \right\} \right] |\bar{D}|^2 
\nonumber \\
&& \qquad \qquad \left. + \frac{\sqrt{6I\epsilon_H}}{2}(-\eta)^{-1}
\sin \theta (\bar{\Gamma}^{'}\bar{D}^{*}+\bar{\Gamma}^{*'}\bar{D})
-\frac{\sqrt{6I\epsilon_H}}{2}(-\eta)^{-2}
\sin \theta (\bar{\Gamma}\bar{D}^{*}+\bar{\Gamma}^{*}\bar{D}) \right] \ ,
\label{eq:lagbegin}
\end{eqnarray}
where we have defined
$\sin \theta \equiv k_y a/ kb$. This $\theta$ represents the direction dependence. 
In the isotropic limit $I=0$, 
the Lagrangian for $\bar{\Gamma}$ becomes the familiar one for gravitational 
waves in a Friedman-Lemaitre universe.

In a similar way, we can derive the quadratic action for physical variables
in the 2-dimensional scalar sector. The details can be found in the Appendix A. 
Moreover, it is straightforward to deduce the action in the slow roll approximation.
The resultant action is given by
\begin{eqnarray}
S^{\rm  scalar} &=& \int d\eta d^3k 
\left[ L^{GG}+L^{JJ}+L^{\phi\phi}+L^{\phi G}
+L^{\phi J}+L^{JG} \right] \ , 
\label{sca-action} \\
L^{GG}&=&\frac{1}{2}|\bar{G}^{'}|^2 
+\frac{1}{2} \left[-k^2+(-\eta) ^{-2} \left\{ 2+3\epsilon _H+3I\epsilon _H
 + 3 I\epsilon _H \sin^{2} \theta \right\} \right] |\bar{G}|^2,
\label{GG} \\
L^{JJ}&=& \frac{1}{2}|\bar{J}^{'}|^2 
+\frac{1}{2} \left[ -k^2 +(-\eta )^{-2} \left\{ 2+9\epsilon _H -3\eta _H 
          - 6 I\epsilon _H \sin^2 \theta \right\} \right] |\bar{J}|^2, 
\label{JJ} \\
L^{\phi \phi} &=& \frac{1}{2} | \delta\bar{ \phi}^{'}|^2
 +\frac{1}{2} \left[ -k^2 +(-\eta) ^{-2} \left\{ 2 + 9\epsilon _H 
 -\frac{3\eta _H}{1-I}-\frac{12I}{1-I}+\left( 12I \epsilon _H +\frac{24I}{1-I} \right)
  \sin^2 \theta \right\} \right] | \delta\bar{\phi}|^2, 
\label{dphi} \\
L^{\phi G} &=& -3 I \sqrt{\frac{\epsilon _H}{1-I}} 
(-\eta )^{-2} \sin^2\theta
\left(\bar{G} \delta\bar{\phi}^* +\bar{G}^* \delta\bar{\phi} \right) \ , 
\label{phi:G} \\
L^{\phi J} &=& \sqrt{ \frac{6I}{1-I}} (-\eta )^{-1} \sin\theta 
\left(\delta\bar{ \phi}^{*'}\bar{J}+\delta\bar{\phi}^{'}\bar{J}^{*}\right)
- \sqrt{ \frac{6I}{1-I}}(-\eta )^{-2}\sin \theta
\left(\delta\bar{ \phi}^{*} \bar{J}+\delta\bar{\phi}\bar{J}^*\right) 
\ , 
\label{phi:J} \\
L ^{JG} &=& -\frac{\sqrt{6I \epsilon _H}}{2} (-\eta ) ^{-1} \sin\theta
\left(\bar{G}^{*'}\bar{J} + \bar{G}^{'} \bar{J}^{*}\right)
+ \frac{\sqrt{6I \epsilon _H}}{2} (-\eta )^{-2}\sin\theta
\left(\bar{G}^{*}\bar{J}+\bar{G}\bar{J}^*\right) \ , 
\label{J:G}
\end{eqnarray}
where we defined canonical variables
\begin{equation}
\bar{G}\equiv\sqrt{2}bG \ , \quad  
\bar{J}\equiv \frac{f |k_x| }{k}J \ , \quad  
\delta\bar{ \phi} \equiv b \delta \phi      \ .
\end{equation}
Here, $\bar{G} \ , \bar{J}$ and $\delta\bar{\phi}$ represent
the gravitational waves, the vector waves, and the scalar perturbations, respectively.
The above action shows there exist the interaction among these variables.
We notice the scalar part (\ref{dphi})
contains $I$ without suppression by a slow-roll parameter $\epsilon _H$. 
Therefore, to obtain the quasi-scale invariant spectrum of curvature perturbation, $I$ itself has to be small.

From the actions (\ref{eq:lagbegin}) and (\ref{sca-action}), 
we see there are two sources of statistical anisotropy of fluctuations. 
First, the statistical anisotropy of fluctuations comes from the anisotropic 
expansion itself. Intuitively, this can be understood from the anisotropic 
effective Hawking temperature $H_{\rm eff}/2\pi$, where $H_{\rm eff}$
denotes the effective expansion rate. 
Indeed, the expansion rate in the direction of the background vector is relatively small,
hence the effective Hawking temperature is low.  Then, this direction has less 
fluctuation power compared to the other directions.
Thus, the effective Hawking temperature 
induces the anisotropy in the power spectrum of fluctuations.
This effect is encoded in (\ref{GG}), (\ref{JJ}), and (\ref{dphi}).
The other source of the statistical anisotropy of fluctuations comes from the couplings
(\ref{phi:G}), (\ref{phi:J}) and (\ref{J:G}) due to the background vector field. 
The essential structure of couplings can be understood without complicated calculations.
Take a look at the following term
\begin{eqnarray}
  \sqrt{-g} g^{\mu\alpha} g^{\nu\beta} f^2(\phi) F_{\mu\nu} F_{\alpha\beta} \ .
\end{eqnarray}
Here, we should recall the order of magnitude of background quantities
$$
 \frac{f^2 v^{\prime 2}}{a^2} \sim I \epsilon_H \ , \quad
 \frac{f_\phi}{f} \sim \frac{\kappa^2 V}{V_\phi}
 \sim \frac{1}{\sqrt{\epsilon_H}} \ .
$$
For example, to obtain the $J-G$ coupling, one of $F_{\mu\nu}$ have to be replaced
by the background quantity $v'$. Hence, the coefficients in the $J-G$ coupling
should be proportional to $f v'$ which is of the order of $\sqrt{I\epsilon_H}$. 
This explains the strength of the coupling in (\ref{J:G}). 
Similarly, $J-\delta\phi$ coupling should be proportional to $f_\phi v'$
 because we have to take the variation with respect to $\phi$.
 Hence, we can estimate its magnitude to be $\sqrt{I}$. 
  This agrees to the interaction term (\ref{phi:J}). Finally, the coupling
$G-\delta\phi$ has a magnitude of the order of $f_\phi v^{\prime 2}$
which is proportional to $I \sqrt{\epsilon_H}$.
This shows a good agreement with the coupling (\ref{phi:G}).
Thus, we can understand why there is a hierarchy among the couplings of
the gravitational waves, the vector waves and the scalar field.

\subsection{Numerical results}

In this section, we will calculate power spectrum of various variables.
To set the initial conditions, we need to quantize this system 
by promoting canonical variables to operators
which satisfy the following canonical commutation relations
\begin{eqnarray}
\left[ Q_a(\eta, {\bf x}), P_b(\eta, {\bf y}) \right] 
= i \delta _{ab} \delta({\bf x}-{\bf y}) \ , \quad
\left[ Q_a(\eta, {\bf x}), Q_b(\eta, {\bf y}) \right]
= \left[ P_a(\eta, {\bf x}), P_b(\eta, {\bf y}) \right] =0
\end{eqnarray}
where $Q_a$ $(1\leq a \leq 5)$ denote the variables 
$\bar{\Gamma},\bar{D},\delta\bar{\phi},\bar{G},\bar{J}$ in this order, 
while $P_a$ is their conjugate momentum defined by 
$P_a \equiv \delta L / \delta Q'_a $.
The point is that, with a given wave number, the actions (\ref{eq:lagbegin}) 
and (\ref{sca-action}) reduce to 
  those of independent harmonic oscillators in the subhorizon limit 
  $-k\eta \gg 1$  
\begin{equation}
S = \sum_{a=1}^{5} \frac{1}{2}\int d\eta d^3 k \left[ |Q^{'}_a|^2-k^2|Q_a|^2 \right] \ .
\end{equation}
We choose the Bunch-Davis vacuum state $|0 \rangle$ 
by imposing the following initial conditions at an initial time $\eta _i$
\begin{eqnarray}
Q_a (\eta_i) = \sqrt{\frac{1}{2k}}(a_{a,\bf k}+a^\dagger _{a,\bf -k} ) \ , \quad
Q'_a (\eta_i) = -i\sqrt{\frac{ k}{2}}(a_{a,\bf k}-a^\dagger _{a,\bf -k}) \ , 
\label{eq:Qini}
\end{eqnarray}
where $a_{a,{\bf k}}$ is an annihilation operator whose commutation relations are given by
\begin{eqnarray}
\left[ a_{a,\bf k} , a^\dagger _{b\bf k^\prime } \right] 
 = \delta_{ab}\delta ^{(3)}({\bf k-k^\prime}), \qquad
\left[ a_{a, \bf k} , a_{b,\bf k^\prime } \right] 
= 0 \ .
\end{eqnarray}
It is easy to verify the commutation relations imposed for $P_a \ , Q_a$. 
The Bunch-Davis vacuum $|0 \rangle$ is defined
by $a_{a,\bf k}|0 \rangle  =0$.

Later, the system  evolves according to the actions (\ref{eq:lagbegin}) and
(\ref{sca-action}) and the variables $Q_a(\eta )$
are given by linear combination of $a_{b,{\bf k}}$ as
\begin{equation}
Q_{a,\bf k}(\eta )
= \sum_{b=1}^{5}
\left[ c_{ab}(\eta ) a_{b,\bf k} +c_{ab}^\ast (\eta ) a_{b,\bf -k}^\dagger \right] \ ,
\end{equation}
where the transfer matrix $c_{ab}$ depends on the wave number 
${\bf k}$ and the time $\eta$.
The time dependence is determined by solving the classical equations 
of motion with the initial conditions given by the coefficients 
in Eqs.(\ref{eq:Qini}). 
And the expectation values of operators in this vacuum state are evaluated as
\begin{equation}
\langle 0 \big| Q_{a \bf k}(\eta)Q^{\ast}_{b \bf p}(\eta) \big| 0 \rangle 
= \sum_{d=1}^{5} c_{ad}c_{bd}^{\ast}  \delta({\bf k} + {\bf p})  \ . 
\end{equation}
Especially, we are interested in the power spectrum of
 the scalar perturbations, cross mode and plus mode of gravitational waves, 
 the cross correlation between scalar
  perturbations and the plus mode of gravitational waves, 
 and the linear polarization of gravitational waves. 
  The power spectrum of scalar perturbations is given by
\begin{eqnarray}
\langle 0 \big|\delta \bar{\phi}_{\bf k}(\eta) 
               \delta \bar{\phi}_{\bf p}(\eta) \big| 0 \rangle
= \sum_{d=3}^{5} |c_{3d}(\eta ) |^2 \delta({\bf k} + {\bf p}) 
\equiv P_{\delta\phi} ({\bf k}) \delta({\bf k} + {\bf p})
    \ ,
\end{eqnarray}
where we took into account the fact that
the 2-dimensional vector sector is decoupled from the scalar sector.
The power spectrum of  the cross and  plus mode of
gravitational waves read 
\begin{eqnarray}
&&  \langle 0 \big| \bar{\Gamma}_{\bf k}(\eta) \bar{\Gamma}_{\bf p}(\eta) \big| 0 \rangle
 = \sum_{d=1}^{2} |c_{1d}(\eta ) |^2 \delta({\bf k} + {\bf p}) 
 \equiv P_{\Gamma} ({\bf k}) \delta({\bf k} + {\bf p}) \ , \\
&&  \langle 0 \big| \bar{G}_{\bf k}(\eta) \bar{G}_{\bf p}(\eta) \big| 0 \rangle
  = \sum_{d=3}^{5} |c_{4 d}(\eta ) |^2 \delta({\bf k} + {\bf p})
 \equiv P_{G} ({\bf k}) \delta({\bf k} + {\bf p})     \ .
\end{eqnarray}
Here, we used the decoupling of the 2-dimensional scalar sector and the vector sector.
We can also calculate the cross correlation between the plus mode
of gravitational waves and the scalar perturbations
\begin{eqnarray}
\langle 0 \big| \delta \bar{\phi}_{\bf k}(\eta) \bar{G}_{\bf p}  (\eta) \big| 0 \rangle
= \sum_{d=3}^{5} c_{3d}(\eta )c_{4d}^{\ast}(\eta ) \delta({\bf k} + {\bf p})
\equiv P_{\delta\phi G} ({\bf k}) \delta({\bf k} + {\bf p})  \ .
\end{eqnarray}
The linear polarization of gravitational waves can be calculated 
once the power spectrum of the cross and plus mode are calculated.

\begin{figure}[htbp]
\includegraphics[width=100mm]{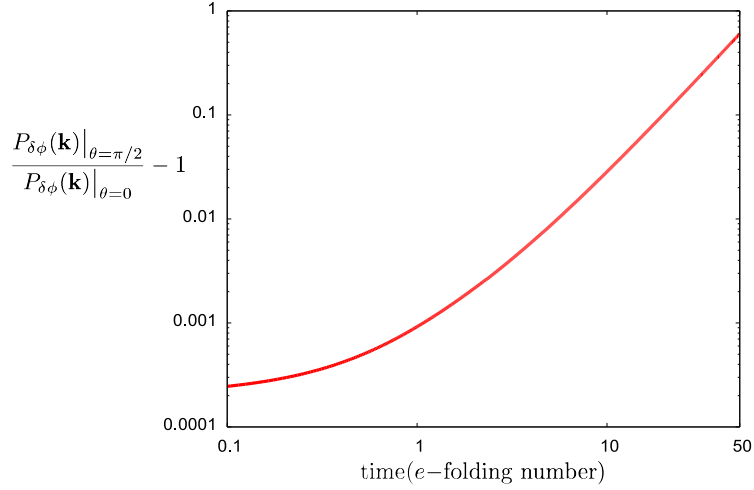}
\caption{Evolution of anisotropy of curvature perturbations.
Here we depicted the anisotropy 
$
P_{\delta\phi} ({\bf k}) \big|_{\theta = \pi/2} /
P_{\delta\phi} ({\bf k}) \big|_{\theta = 0}  - 1
$
 as a function of time. We set $e$-folding number to be zero 
 at the time of horizon crossing of the given mode. The both axes are taken in log scale.}
\label{fg:curv}
\end{figure}
\begin{figure}[htbp]
\includegraphics[width=100mm]{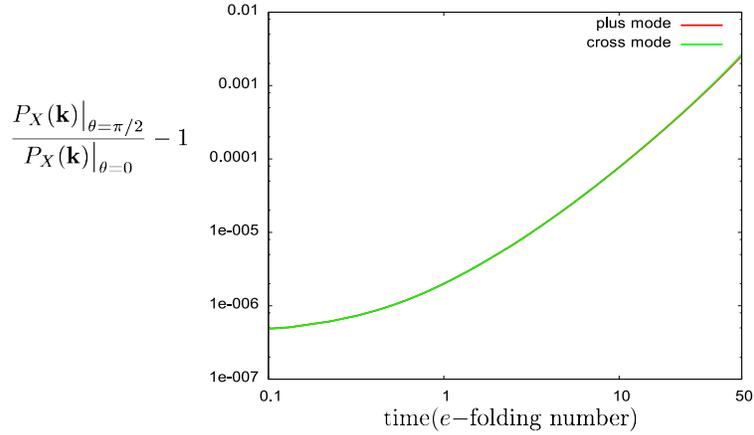}
\caption{Evolution of anisotropy in gravitational waves.
Here we depicted the anisotropy 
$
P_{X} ({\bf k})\big|_{\theta = \pi/2} /P_{X} ({\bf k})\big|_{\theta = 0}  - 1
$, where $X=\Gamma , G$, as a function of time. The axes are in log scale.
As one can see, the difference between two modes is quite small. } 
   \label{fg:gw}
\end{figure}
\begin{figure}[htbp]
\includegraphics[width=100mm]{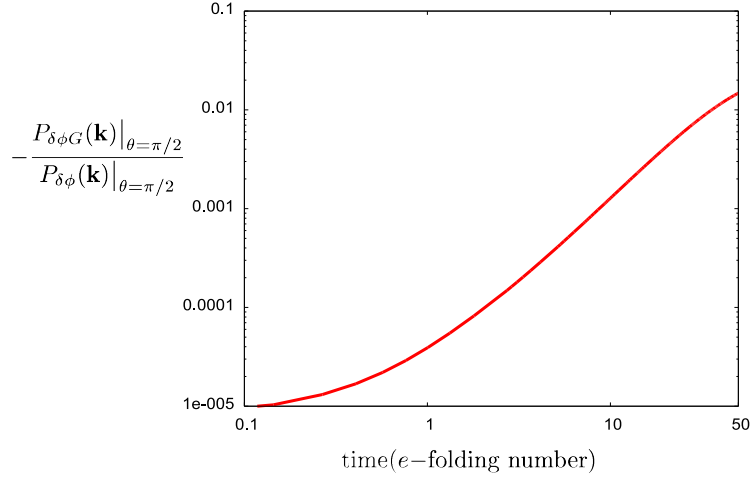}
\caption{Evolution of cross correlation between curvature perturbations 
and the plus mode of gravitational waves. Here we depicted the value 
$
-P_{\delta\phi G} ({\bf k}) \big|_{\theta = \pi/2} / 
P_{\delta\phi} ({\bf k}) \big|_{\theta = \pi/2}
$ as a function of the time. The axes are in log scale.}
   \label{fg:ST}   
\end{figure}
\begin{figure}[htbp]
\includegraphics[width=100mm]{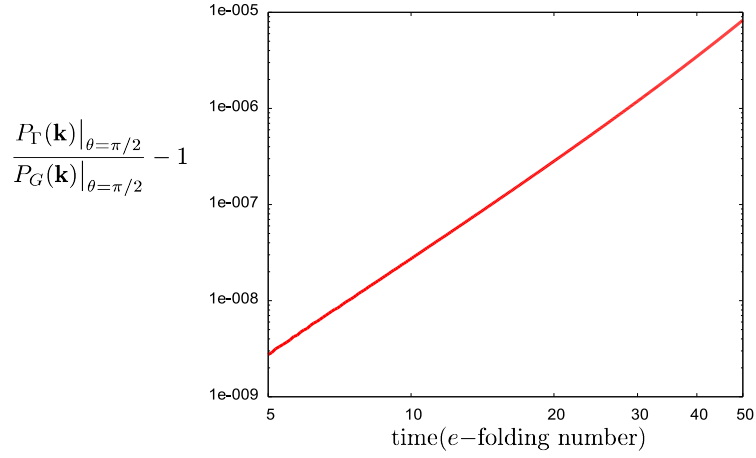}
\caption{Evolution of the linear polarization of gravitational waves.
Here we depicted the value 
$
P_{\Gamma} ({\bf k}) \big|_{\theta = \pi/2}/
P_{G} ({\bf k}) \big|_{\theta = \pi/2} - 1
$ as a function of time. The axes are in log scale.} 
   \label{fg:polar}
\end{figure}

Now, we numerically calculate the transfer matrix $c_{ab}$ with parameters 
$\epsilon_H = 10^{-2},\ I=10^{-5}$ and evaluate various statistical quantities. 
First of all, we need to calculate the statistical anisotropy in the curvature
perturbations. 
We depict the time evolution of the anisotropy in the curvature perturbation  
\begin{eqnarray}
\frac{P_{\delta\phi} ({\bf k}) \big|_{\theta = \pi/2}}{
      P_{\delta\phi} ({\bf k}) \big|_{\theta = 0}} 
                                  - 1
\end{eqnarray}
as a function of time in Fig.\ref{fg:curv}. 
Taking look at Fig.\ref{fg:curv}, we see
the anisotropy grows with the squared of the e-folding number.
The amplitude is larger than the expected one due to the coupling proportional
 to $\sqrt{I}$ which is larger than the expected one $\sqrt{I\epsilon_H}$. 
Similarly, in Fig.\ref{fg:gw}, we show the power spectrum of the cross
and plus mode of gravitational waves
\begin{eqnarray}
\frac{P_{X} ({\bf k}) \big|_{\theta = \pi/2}}{P_{X} ({\bf k}) \big|_{\theta = 0}}
                          - 1 \ ,
\end{eqnarray}
where $X= \Gamma , G $. 
Next, we are interested in the cross correlation between the curvature perturbations
and the gravitational waves.
In Fig.\ref{fg:ST}, we depict the cross correlation between the curvature perturbation and the plus mode of gravitational waves which is normalized 
by the power spectrum of scalar perturbations 
\begin{eqnarray}
\frac{P_{\delta\phi G} ({\bf k}) \big|_{\theta = \pi/2}}
{P_{\delta\phi} ({\bf k}) \big|_{\theta = \pi/2} }
\end{eqnarray}
In Fig.\ref{fg:ST}, we see the cross correlation between
the scalar perturbations and the gravitational waves grows
quadratically in the e-folding number. 
The amplitude is proportional to the coupling between $G$ and $\delta \phi$,
i.e., $I \sqrt{\epsilon_H}$. 
The difference between the power spectrum of two polarization modes
normalized by the power spectrum of the plus mode of gravitational waves
\begin{eqnarray}
\frac{P_{\Gamma} ({\bf k}) \big|_{\theta = \pi/2}}
{P_{G} ({\bf k}) \big|_{\theta = \pi/2}}
                      - 1  
\end{eqnarray}
characterizes the linear polarization of the gravitational wave.
As you can see from  Fig.\ref{fg:gw} and Fig.\ref{fg:polar}, the anisotropy of power 
spectrum grows quadratically in the e-folding number, while the difference between the two polarization modes shows the growth of higher power in e-folding number.

\subsection{Analytical Estimation}

In this section, we treat the anisotropy perturbatively and estimate its magnitude by 
using perturbation in the interaction picture. 
In the interaction picture, 
the expectation value for a physical quantity ${\cal O} (\eta)$ is given by
\begin{equation}
\langle in \left| {\cal O} (\eta) \right |in \rangle 
= \left< 0 \left| 
\left[ \bar{T}\exp \left( i \int ^{\eta}_{\eta_i} H_I(\eta ^{'}) d\eta ^{'} 
\right) \right] {\cal O} (\eta ) 
\left[ T \exp \left( -i\int ^{\eta}_{\eta_i} H_I(\eta ^{'}) d\eta ^{'} \right) \right] 
\right| 0 \right> \ ,
\end{equation}
where $|in \rangle$ is an in vacuum in the interaction picture, 
 $T$ and $\bar{T}$ denote a time-ordered and an anti-time-ordered product
 and $H_I$ denotes  the interaction part of Hamiltonian
 in this picture. 
 This is equivalent to the following
\begin{equation}
\langle in \left| {\cal O} (\eta) \right| in \rangle 
= \sum _{N=0}^{\infty} i^N \int _{\eta_i}^{\eta} d\eta _N\int _{\eta_i}^{\eta_N} 
d\eta _{N-1} \cdots \int _{\eta_i}^{\eta_2} d\eta _{1} 
\left<0 \left| 
\left[ H_I(\eta_1),\left[H_I(\eta_2), \cdots \left[H_I(\eta _N),{\cal O}(\eta) \right] 
\cdots \right] \right] 
\right| 0 \right>.
\end{equation}
The successive approximation is made by truncating the summation at a certain order $N$.
In our analysis, we assume the noninteracting part of Hamiltonian to be that of free fields in deSitter spacetime 
\begin{equation}
L_0 = \sum _{n=1}^{5} \left[ \frac{1}{2} |Q^{'}_n|^2-\frac{1}{2} 
                    \left( k^2 - 2(-\eta )^{-2} \right) |Q_n|^2 \right] \ ,
\end{equation}
and the operators in the interaction picture are given by
\begin{eqnarray}
Q_{n,{\bf k}}(\eta) &=& u(\eta)  a_{n,{\bf k}} 
+ u(\eta)^{*} a_{n,{\bf -k}}^{\dagger}, 
\label{eq:nonint}\\
u(\eta) &\equiv &\sqrt{\frac{1}{2k}}e^{-ik\eta}\left(1-\frac{i}{k\eta} \right). 
\label{eq:modefunc}
\end{eqnarray}
And the rest of the Lagrangian (\ref{eq:lagbegin})-(\ref{J:G})
is regarded as the interaction part $L_I = L^{(2)}-L_0$. 
To see the leading effect on the anisotropy in the scalar perturbation, 
which is of the order of $I$, we evaluate the correction due to the interaction
 given by
\begin{eqnarray}
H_I^{\phi J} &\equiv & \int d^3 k \left[ - L^{\phi J} \right] \nonumber\\
&=& \int d^3k \left[ -\sqrt{ \frac{6I}{1-I}} (-\eta )^{-1} \sin\theta 
\left(\bar{\delta \phi}^{\dagger'}\bar{J} +\bar{\delta\phi}^{'}\bar{J}^{\dagger}\right)
+ \sqrt{ \frac{6I}{1-I}}(-\eta )^{-2}\sin \theta
\left(\bar{\delta \phi}^{\dagger} \bar{J} +\bar{\delta\phi} \bar{J}^{\dagger}\right)
\right] 
\ .
\end{eqnarray}
 Note that in the analogy with the slow-roll parameter in the ordinary slow-roll inflation, the term proportional to 
 $I\sin^2 \theta \delta\bar{\phi}
 \delta\bar{\phi}^{\dagger}$ in (\ref{dphi}) can be expected to give the anisotropy  
$\delta \langle in \left| \delta \bar{\phi}_{\bf k}
                         \delta \bar{\phi}_{\bf p} \right| in \rangle 
/ \langle 0 \left| \delta \bar{\phi}_{\bf k} \delta \bar{\phi}_{\bf p} \right| 0 \rangle 
\sim \sin ^2 \theta I N(k) $ 
where $N(k)$ is the $e$-folding number from the horizon exit. 
Thus, the leading correction comes from the interaction
through the term $H^{\phi J}_{I}$. The leading correction is given by
\begin{equation}
\delta \langle in \left| \delta \bar{\phi}_{\bf k} (\eta) 
                   \delta \bar{\phi}_{\bf p} (\eta) \right| in \rangle  
= i^2 \int_{\eta _i}^{\eta}d\eta_2\int_{\eta _i}^{\eta _2} d\eta _1 
\left< 0 \left|
 \left[ H^{\phi J}_I(\eta_1),\left[H^{\phi J}_I(\eta_2),\delta \bar{\phi}_{\bf k} (\eta)
 \delta \bar{\phi}_{\bf p} (\eta) \right] \right] 
 \right| 0 \right>.
\end{equation}
Using Eqs.(\ref{eq:nonint}) and commutation relations for the creation and annihilation operators, we obtain the anisotropy expressed as follows
\begin{eqnarray}
\frac{\delta \langle in \left| \delta \bar{\phi}_{\bf k}
                             \delta \bar{\phi}_{\bf p} \right| in \rangle}
{\langle 0 \left| \delta \bar{\phi}_{\bf k} 
                 \delta \bar{\phi}_{\bf p} \right| 0 \rangle} (\eta) 
&=& \frac{24 I}{1-I} \sin ^2 \theta 
 \int ^{\eta} _{\eta_i} d\eta _2 \int ^{\eta_2}_{\eta_i} d\eta_1 
  \frac{8}{|u(\eta)|^2} {\rm Im}
\left[  -(-\eta_2)^{-1}  u^{'}(\eta_2) u^{*}(\eta) 
   + (-\eta _2 )^{-2} u(\eta_2 )u^{*}(\eta )  \right] \nonumber\\
&\ &\, \times {\rm Im}
\left[ u(\eta_1 )u^{*}(\eta _2)\left\{ -(-\eta_ 1)^{-1} u^{'}(\eta _1)u^{*}(\eta ) 
 + (-\eta_1)^{-2} u(\eta_1 )u^{*}(\eta) \right\}  \right] \ ,
\end{eqnarray} 
where ${\rm Im}$ denotes the imaginary part.
Substituting the function form of $u$ (\ref{eq:modefunc}) 
and introducing time variables $\chi \equiv k\eta$, $\chi_1 \equiv k\eta_1$ and
$\chi_2 \equiv k\eta_2$, we have
\begin{eqnarray}
\frac{\delta \langle in \left| \delta \bar{\phi}_{\bf k}
                             \delta \bar{\phi}_{\bf p} \right| in \rangle}
{\langle 0 \left| \delta \bar{\phi}_{\bf k}
                             \delta \bar{\phi}_{\bf p} \right| 0 \rangle} (\chi) 
&=& \frac{6I}{1-I} \sin ^2 \theta \int ^{\chi}_{\chi_i}d\chi_2 \int ^{\chi_2}_{\chi_i} d\chi_1 
 \frac{8}{1+\frac{1}{(-\chi)^2}}\frac{1}{-\chi_1}\frac{1}{-\chi_2}\left[ \cos(-\chi_2+\chi)-\sin(-\chi_2+\chi) \frac{1}{\chi} \right] \nonumber\\
&\ &\, \times 
\bigg[ \cos(-2\chi_1+\chi+\chi_2) \left( 1+\frac{1}{\chi\chi_1}-\frac{1}{\chi\chi_2}+\frac{1}{\chi_1\chi_2} \right) \nonumber\\ 
&\ &\, + \sin(-2\chi_1+\chi+\chi_2)\left( -\frac{1}{\chi\chi_1\chi_2}+\frac{1}{\chi_1}-\frac{1}{\chi}-\frac{1}{\chi_2} \right) \bigg].\label{eq:integrand}
\end{eqnarray}
The contribution to the integral 
from the subhorizon $- \chi_1 \gg 1$ is negligible. 
 In the limit of superhorizon $-\chi_1 \ll 1$, we also have
 $-\chi_2 \ll 1 , -\chi \ll 1$. Hence,  the integrand in Eq.(\ref{eq:integrand}) approximately becomes $8/\chi_1 \chi_2 $.
  Thus, the anisotropy can be evaluated as
\begin{eqnarray}
\frac{\delta \langle in \left| \delta \bar{\phi}_{\bf k}
                             \delta \bar{\phi}_{\bf p} \right| in \rangle}
{\langle 0 \left| \delta \bar{\phi}_{\bf k}
                             \delta \bar{\phi}_{\bf p} \right| 0 \rangle} (\chi) 
&=& \frac{6I}{1-I} \sin ^2 \theta \int^{\chi}_{-1} d\chi_2 \int^{\chi_2}_{-1} d\chi_1 \frac{8}{\chi_1 \chi_2} \nonumber\\
&=& \frac{24I}{1-I} \sin ^2 \theta \ N^2(k),
\end{eqnarray}
where $N(k) \equiv -\ln (-k\eta)$ is the $e$-folding number from the horizon exit.

For the anisotropy of both two polarizations of gravitational waves, 
the similar calculations give
\begin{equation}
\frac{\delta \langle in \left|  \bar{\Gamma}_{\bf k} 
                                \bar{\Gamma}_{\bf p} \right| in \rangle}
{\langle 0 \left|  \bar{\Gamma}_{\bf k} 
                                \bar{\Gamma}_{\bf p} \right| 0 \rangle}
 = \frac{\delta \langle in \left| \bar{G}_{\bf k} \bar{G}_{\bf p} \right| in \rangle}
 {\langle 0 \left| \bar{G}_{\bf k} \bar{G}_{\bf p} \right| 0 \rangle} 
 = 6 I \epsilon_H \sin ^2 \theta \ N^2(k)  \ ,
\end{equation}
where we used the interaction term in the action (\ref{eq:lagbegin})
for $\bar{\Gamma}$ and that in (\ref{J:G}) for $\bar{G}$. 
It is interesting to calculate the cross correlation. 
The leading contribution comes from $H_I^{JG}$ and $H_I^{\phi J}$.
The result is as follows:
\begin{eqnarray}
\frac{\langle in \big| \delta\bar{\phi}_{\bf k} \bar{G}_{\bf p} \big| in \rangle}
{  \langle 0\big| \delta\bar{\phi}_{\bf k} \delta\bar{\phi}_{\bf p} \big| 0 \rangle}
\simeq - 24 I \sqrt{\frac{\epsilon_H}{1-I}}  N^2(k)  \ .
\label{cross}
\end{eqnarray}
As we will soon see, this might give a detectable number. 
On the other hand, 
it turns out that the linear polarization of the gravitational wave 
must be the effect of the order higher than $I$.

\section{Cosmological Implication}

Now, we are in a position to discuss cosmological implication
of an anisotropic inflationary scenario. As we have listed up in section I,
there are many interesting phenomenology in the anisotropic inflation. 
Here, we summarize our results:
\begin{itemize}
\item There exists statistical anisotropy in curvature perturbations
of the order of $24 I N^2(k)$.
\item There exists statistical anisotropy in gravitational waves
of the order of $6I \epsilon_H N^2(k)$.
\item These exists the cross correlation 
between curvature perturbations and gravitational waves
of the order of $-24 I \sqrt{\epsilon_H} N^2(k)$. 
\item There is small linear polarization of gravitational waves.
\end{itemize} 
Due to the interaction on superhorizon scales, there is an
enhancement factor $N^2(k)$ in the above quantities. 
Because of this enhancement, even when the anisotropy of the spacetime
is quite small, say $\Sigma/H \sim 10^{-7}$ in our example, 
the statistical anisotropy imprinted in primordial fluctuations
can not be negligible in  precision cosmology.   

Let us be more precise. 
The anisotropy is often parameterized by
\begin{eqnarray}
  P({\bf k} ) = P(k) \left[ 1 + g_* \sin^2 \theta \right] \ .
\end{eqnarray}
We have estimated $g_* \simeq 24 I N^2 (k)$ which could be large.
The current observational limit of the statistical anisotropy 
for the curvature perturbations is characterized by 
$g_* < 0.3$~\cite{Pullen:2007tu}. Let us suppose $g_* =0.2$.
Then, we have the anisotropy in the gravitational waves with
$g_* \simeq 10^{-3}$. The linear polarization is of the order of $10^{-6}$.
Using the definition of curvature perturbations 
$\zeta = \delta\bar{\phi}/\sqrt{2 \epsilon_H}$, one can translate
the cross correlation (\ref{cross}) between the scalar perturbations and
gravitational waves to that between the curvature perturbations and
gravitational waves normalized by the power spectrum of curvature perturbations:
\begin{eqnarray}
\frac{\langle in \big| \zeta_{\bf k} \bar{G}_{\bf p} \big| in \rangle}
{  \langle 0\big| \zeta_{\bf k} \zeta_{\bf p} \big| 0 \rangle}
\simeq - 20 I N^2(k) \epsilon_H 
\sim -  g_* \epsilon_H \sim -  2 \times 10^{-3} \ ,
\end{eqnarray}
where we used $g_{*} \sim 0.2 $ and $\epsilon_H \sim 10^{-2}$.
Since the current constraints on the $TB/TE$ ratio is of the order 
of $10^{-2}$~\cite{Komatsu:2010fb}, we need to improve
the accuracy by one more order,
which might be achieved by PLANCK. 

In \cite{Gumrukcuoglu:2010yc}, it is pointed out that the sign of
$g_*$ is different from the observed one. However, it might be possible to
modify the model so that the sign of $g_*$ is flipped. 
For example, we can consider two vector fields.
 Then, the orthogonal direction to the plane determined by two vectors 
 becomes a preferred direction. In this case, we can expect the sign of
 $g_{*}$ becomes opposite. We can also utilize anti-symmetric tensor fields
to achieve the same aim. The details will be reported elsewhere.  

\section{Conclusion}

We have studied the statistical nature of primordial fluctuations 
from an anisotropic inflation which is realized by a vector field 
coupled to an inflaton.
First, we have classified metric fluctuations according to the 2-dimensional
rotational symmetry. To choose a convenient gauge in an anisotropic universe,
 we have started from the flat slicing gauge in an isotropic universe
 and made an appropriate gauge transformation to get a canonical gauge
 in an anisotropic universe. This gauge choice has made the subsequent analysis
 and the interpretation of the variables easier.
Using the canonical gauge, we have revealed the structure of the couplings
between curvature perturbations, vector waves, and gravitational
waves. We found that there are two sources for anisotropy, i.e. the anisotropy
due to the anisotropic expansion of the universe and that due to
the anisotropic couplings among variables. 
It turned out that the latter effect is dominant. 
We have numerically obtained power spectra. We also presented
analytical formula using in-in formalism.
Since the coupling between
the curvature perturbations and vector waves is the strongest one,
the anisotropy in the curvature perturbations is larger than
that in  gravitational waves. 
More interestingly, we found the cross correlation between
curvature perturbations and  gravitational waves
which is peculiar to anisotropic inflation.
We also found the linear polarization of gravitational waves.  
Although there are several mechanism to produce circular polarization
in the primordial gravitational waves~\cite{Lue:1998mq}, 
this is the first example which
realized the linear polarization in the primordial gravitational waves.

We have only considered power spectrum for simplicity.
However, as is pointed out in the paper~\cite{Yokoyama:2008xw}, 
the statistical anisotropy
could appear in the non-Gaussianity strongly and modify the 
shape of the bispectrum and trispectrum. Hence, it is interesting to
study non-Gaussianity in anisotropic inflation models. 

We can extend anisotropic inflation in various ways.
Although we have investigated a chaotic inflation in this paper,
it is easy to extend the analysis to other inflation models. 
It is possible to incorporate multi-vector fields. From the string theory
point of view, it is intriguing to consider anti-symmetric tensor field.
It is also interesting to consider other Bianchi type models~\cite{Dechant:2008pb}
in the context of anisotropic inflation.

\begin{acknowledgements}
SK would like to thank the YITP members in Kyoto for warm hospitality. 
A part of this work was done while SK was visiting YITP supported 
by JSPS Grant-in-Aid for Scientific Research (A) 21244033. SK is 
supported by an STFC rolling grant.
JS is supported by the Japan-U.K. Research Cooperative Program, 
Grant-in-Aid for  Scientific Research Fund of the Ministry of 
Education, Science and Culture of Japan No.18540262,
Grant-in-Aid for  Scientific Research on Innovative Area No.21111006
and the Grant-in-Aid for the Global COE Program 
``The Next Generation of Physics, Spun from Universality and Emergence".
\end{acknowledgements}

\appendix

\section{derivation of reduced action for 2d scalar sector}

In this appendix, we present a detailed derivation of the action
for the 2-dimensional scalar sector. Here, we need to eliminate non-dynamical fields
$\delta A_0$, $\beta_1$, $\beta_2$ and $\Phi$.

First, let us see the terms concerning $\delta A_0$ in Fourier space
\begin{eqnarray}
\frac{b^2}{2a^2}f^2k^2| \delta A_{0} |^2 
+\delta A_{0}^* \left[ 
-ik_x\frac{b^2}{2a^2}f^2v^{'}(G+\Phi)
+ik_y\frac{f^2}{2}J^{'}
+ik_x\frac{b^2}{a^2}ff_{\phi}v^{'}\delta\phi \right] 
+ {\rm c.c.}\ . \nonumber \\   
\end{eqnarray}
By completing the square as
\begin{eqnarray}
\frac{b^2}{2a^2}f^2k^2~ \bigg|~ \delta A_{0} 
-\frac{ik_xv^{'}}{k^2} \left( G+\Phi-2\frac{f_\phi}{f}\delta\phi\right)
+i\frac{k_y}{k^2}\frac{a^2}{b^2}J^{'} ~\bigg| ^2 
-\frac{b^2}{2a^2}\frac{f^2}{k^2}
 \left| k_xv^{'} \left( G+\Phi-2\frac{f_\phi}{f}\delta\phi\right)
-\frac{a^2}{b^2}k_yJ^{'} \right| ^2 \ , \label{eq:A0}
\end{eqnarray}
the variable $\delta A_0$ can be eliminated because the first squared term 
vanishes after substituting the equation of motion 
for $\delta A_0$. 
Similarly, the terms containing $\beta_1$ are given by
\begin{eqnarray}
\frac{1}{4}k_y^2|\beta_{1}|^2 
 +\beta_1^{*} \left[ -\frac{1}{4}k_xk_y\beta_{2}+ik_x\frac{b^{'}b}{a}\Phi
-ik_x\frac{b^2\phi ^{'}}{2a} \delta\phi \right] + {\rm c.c.}\ . 
\end{eqnarray}
Completing the square gives 
\begin{eqnarray}
 \frac{k_y^2}{4} \left|~ \beta_1+ \frac{k_x}{k_y}\beta_{2}
+4i\frac{k_x}{k_y^2}\frac{bb^{'}}{a}\Phi 
-2i\frac{k_x}{k_y^2}\frac{b^2\phi ^{'}}{a} \delta\phi ~\right|^2 
-\frac{k_x^2}{4} \left|~ \beta_{2}-4i\frac{1}{k_y}\frac{bb^{'}}{a}\Phi
+2i\frac{1}{k_y}\frac{b^2\phi ^{'}}{a}\delta\phi ~\right| ^2  \ . \label{eq:B}
\end{eqnarray}
Again, we can ignore the first term.
Taking a look at the terms related to the variable $\beta_2$,
we see only the linear term in $\beta_2$ appears as 
\begin{eqnarray}
\beta_{2}^{*}\bigg[~ \frac{a}{2}\left(\frac{a^\prime}{a}+\frac{b^\prime}{b}
+2\frac{b^{'}}{b}\frac{b^2}{a^2}\frac{k_x^2}{k_y^2} \right) ik_y\Phi 
-\frac{a}{2}\left( \frac{a^{'}}{a}-\frac{b^{'}}{b} \right) ik_yG 
-\frac{a}{2}\phi^\prime \frac{k^2}{k_y^2}\frac{b^2}{a^2}  
ik_y\delta\phi
+\frac{f^2v^{'}}{2a}ik_xJ ~\bigg] + {\rm c.c.} \ . \nonumber\\
\end{eqnarray}
The variation with respect to $\beta_2$ gives
\begin{equation}
\Phi = \frac{1}{\lambda}\left[
\left(\frac{a^\prime}{a}-\frac{b^\prime}{b}\right)G
+\frac{k^2 b^2}{k_y^2a^2}  \phi ^{'} \delta\phi 
-\frac{k_x}{k_y}\frac{f^2v^{'}}{a^2} J \right] \ .  \label{eq:C}
\end{equation}
Here we have defined 
\begin{eqnarray}
\lambda 
= \frac{a^\prime}{a}+\frac{b^\prime}{b}
+2\frac{k_x^2}{k_y^2}\frac{b^2}{a^2}\frac{b^\prime}{b} \ .
\end{eqnarray}
After substituting this result into the action, we obtain the action
for physical variables:
\begin{eqnarray}
S^{(2){\rm scalar}}  =  \int d\eta d^3k \left[ 
 L^{GG}+L^{\phi\phi}+L^{JJ}+L^{G\phi}+L^{\phi J}+L^{JG} \right] \ .
\end{eqnarray}
The term $L^{GG}$ is given by
\begin{eqnarray}
L^{GG} &=&b^2|G^{'}|^2-b^2k^2|G|^2
+\frac{k_y^2}{2k^2}f^2v^{\prime 2}|G|^2 
-\frac{b^2}{2a^2}\frac{f^2v^{\prime2}}{\lambda^2}\frac{k_x^2}{k^2}
\left(\frac{a^\prime}{a}-\frac{b^\prime}{b}\right)^2|G|^2
\nonumber\\
 &\ &
-\frac{4}{\lambda^2}\frac{b^2b^{\prime2}}{a^2}\frac{k_x^2}{k_y^2}
\left(\frac{a^\prime}{a}-\frac{b^\prime}{b}\right)^2|G|^2
-\frac{a^2b^2}{\lambda^2}V
\left(\frac{a^\prime}{a}-\frac{b^\prime}{b}\right)^2|G|^2
-\frac{b^2}{a^2}\frac{f^2v^{\prime2}}{\lambda}\frac{k_x^2}{k^2}
\left(\frac{a^\prime}{a}-\frac{b^\prime}{b}\right)|G|^2
\nonumber\\
 &\ &
 -\frac{2a^2b^2}{\lambda}V
\left(\frac{a^\prime}{a}-\frac{b^\prime}{b}\right)|G|^2
+\frac{bb^{\prime\prime}+b^{\prime2}}{\lambda}
\left(\frac{a^\prime}{a}-\frac{b^\prime}{b}\right)|G|^2
-bb^\prime\biggl\{\frac{1}{\lambda}
\left(\frac{a^\prime}{a}-\frac{b^\prime}{b}\right)
\biggr\}^\prime|G|^2
 \ .
\end{eqnarray}
Using the canonically normalized variable $\bar{G} \equiv \sqrt{2}bG$,
 we finally have 
\begin{eqnarray}
L^{GG} &=& \frac{1}{2}|\bar{G}^{'}|^2 
+\frac{1}{2} \bigg[~\frac{b^{''}}{b}-k^2
+\frac{k_y^2}{2k^2}\frac{f^2v^{'2}}{b^2}
-\frac{1}{\lambda^2} \left( 
a^2 V+\frac{f^2v^{'2}}{2a^2}\frac{k_x^2}{k^2} 
+4\frac{k_x^2}{k_y^2}\frac{b^{'2}}{a^2}\right) 
\left(\frac{a^\prime}{a}-\frac{b^\prime}{b}\right)^2 \nonumber\\
&\ & \hspace{1.9cm} 
-\frac{1}{\lambda}\left(
2a^2V+\frac{f^2v^{'2}}{a^2}\frac{k_x^2}{k^2}
-\frac{b^{\prime\prime}}{b}-\frac{b^{\prime2}}{b^2}\right)
\left(\frac{a^\prime}{a}-\frac{b^\prime}{b}\right)
-\frac{b^\prime}{b}\left\{ \frac{1}{\lambda}
\left(\frac{a^\prime}{a}-\frac{b^\prime}{b}\right)
 \right\}^{\prime} 
~\bigg] |\bar{G}|^2.
\end{eqnarray}
The term $L^{\phi\phi}$ reads
\begin{eqnarray}
L^{\phi\phi} &=& 
\frac{b^2}{2}|\delta\phi^{'}|^2
-\frac{b^2}{2}k^2|\delta\phi|^2
-\frac{a^2b^2}{2}V_{\phi\phi}|\delta\phi|^2 \nonumber\\
 &\ &-2\frac{b^2}{a^2}f_{\phi}^2v^{'2}\frac{k_x^2}{k^2} |\delta\phi|^2 
+\frac{b^2}{2a^2}(ff_{\phi\phi}+f_{\phi}^2)v^{'2}|\delta\phi|^2 
-\frac{b^4}{a^2}\frac{k_x^2}{k_y^2}\phi ^{'2}|\delta\phi|^2 
\nonumber\\
 &\ &-\frac{\phi ^{'2}}{\lambda^2} 
  \left(a^2b^2 V+\frac{b^2f^2v^{'2}}{2a^2}\frac{k_x^2}{k^2} 
+4\frac{k_x^2}{k_y^2}\frac{b^{'2}b^2}{a^2}\right) 
\left(\frac{k^2 b^2}{k_y^2 a^2} \right)^2|\delta\phi|^2 
\nonumber\\
 &\ &+ \frac{\phi ^{'}}{\lambda}
   \left( -a^2b^2 V_{\phi}-ff_{\phi}\frac{b^2v^{'2}}{a^2}
  +4\frac{k_x^2}{k_y^2}\frac{b^3b^{'}}{a^2}\phi ^{'}
  +2\frac{k_x^2b^2}{k^2a^2}ff_{\phi}v^{'2} \right) 
\frac{k^2 b^2}{k_y^2a^2} |\delta\phi|^2 
+\frac{1}{2}
\left\{\frac{b^2\phi^{\prime2}}{\lambda}\frac{k^2b^2}{k_y^2a^2} 
\right\}^\prime|\delta\phi|^2  \ .
\end{eqnarray}
Using the canonically normalized variable 
$\bar{\delta\phi} \equiv b\delta\phi $, we can write
\begin{eqnarray}
L^{\phi\phi} 
&=& \frac{1}{2}|\bar{\delta\phi}^{\prime}|^2 
+\frac{1}{2} \bigg[ -k^2+\frac{b^{''}}{b}-a^2 V_{\phi\phi}
 -4\frac{f_{\phi}^2v^{'2}}{a^2}\frac{k_x^2}{k^2}+(ff_{\phi\phi}
 +f_{\phi}^2)\frac{v^{'2}}{a^2} 
  -2\frac{k_x^2}{k_y^2}\frac{b^2}{a^2}\phi ^{'2} \nonumber\\
 &\ &\hspace{3em} 
-2\frac{\phi^{\prime2}}{\lambda^2}
 \left( a^2 V+\frac{f^2v^{'2}}{2a^2}\frac{k_x^2}{k^2}
  +4\frac{k_x^2}{k_y^2}\frac{b^{'2}}{a^2}\right) 
\left(\frac{k^2 b^2}{k_y^2a^2} \right)^2  
\nonumber\\
 &\ &\hspace{3em} 
 +2\frac{  \phi ^{'}}{\lambda} 
 \left( -a^2 V_{\phi}-\frac{ff_{\phi}v^{'2}}{a^2}
 +4\frac{k_x^2}{k_y^2}\frac{bb^\prime}{a^2}\phi ^{'}
 +2\frac{k_x^2}{k^2}\frac{ff_{\phi}v^{'2}}{a^2} \right) 
\frac{k^2 b^2}{k_y^2a^2} 
 +\frac{1}{2b^2}\left\{\frac{b^2\phi^{\prime2}}{\lambda}
\frac{k^2 b^2}{k_y^2a^2}  
  \right\}^{\prime}~ \bigg]|\bar{\delta\phi}|^2 \ .
\end{eqnarray}
Similarly, the term $L^{JJ}$ reads
\begin{eqnarray}
L^{JJ} &=& \frac{f^2}{2}\frac{k_x^2}{k^2}|J^{\prime}|^2
-\frac{f^2}{2}k_x^2|J|^2 
\nonumber\\
 &\ &
-\left\{a^2 b^2 V+\frac{f^2b^2v^{'2}}{2a^2}\frac{k_x^2}{k^2} 
 +4\frac{k_x^2}{k_y^2}\frac{b^{'2}}{a^2}b^2
-\frac{\lambda a^2}{2}\frac{k_y^2}{k^2}
\left(
2\frac{a^\prime}{a}-4\frac{b^\prime}{b}-2\frac{k^\prime}{k}
-\frac{\lambda^\prime}{\lambda}\right)
\right\}
\left(\frac{v^{\prime}}{\lambda}\frac{f^2}{a^2}\frac{k_x}{k_y}\right)^2
|J|^2 
  \ .
\end{eqnarray}
Using the canonically normalized variable $ \bar{J} \equiv \frac{f|k_x|}{k}J $ ,
we obtain 
\begin{eqnarray}
L^{JJ} &=& \frac{1}{2} |\bar{J}^{\prime}|^2 
-\frac{1}{2} 
\bigg[~ k^2
+\left\{\left(\log\frac{k}{f}\right)^{\prime\prime}
-\left(\log\frac{k}{f}\right)^2 \right\}\nonumber\\
 &\ & 
 +2\left\{a^2b^2 V+\frac{b^2}{2a^2}\frac{k_x^2}{k^2}f^2v^{'2} 
 +4\frac{k_x^2}{k_y^2}\frac{b^{\prime2}b^2}{a^2}
-\frac{\lambda a^2}{2}\frac{k_y^2}{k^2}\left(
2\frac{a^\prime}{a}-4\frac{b^\prime}{b}-2\frac{k^\prime}{k}
-\frac{\lambda^\prime}{\lambda}\right)
\left(\frac{v^{\prime}}{\lambda}\frac{f^2}{a^2}\frac{k_x}{k_y}\right)^2
\right\}
~\bigg]|\bar{J}|^2 
\ .
\end{eqnarray}
The mixed term $L^{G\phi}$ is given by
\begin{eqnarray}
L^{G\phi} &=& 
\left[
-\frac{b^4}{2a^4}f^2v^{\prime2}\frac{k_x^2}{k_y^2}\frac{\phi^\prime}{\lambda}
+2bb^\prime\frac{b^2}{a^2}\frac{k_x^2}{k_y^2}\frac{\phi^\prime}{\lambda}
\left(\frac{a^\prime}{a}-\frac{b^\prime}{b}\right)\right.
\nonumber\\
&\ & 
-\frac{\phi^\prime}{\lambda^2}\left(2a^2b^2 V
+\frac{f^2b^2v^{'2}}{2a^2}\frac{k_x^2}{k^2} 
+4\frac{k_x^2}{k_y^2}\frac{b^{\prime2}b^2}{a^2}\right) 
\left(\frac{a^\prime}{a}-\frac{b^\prime}{b}\right)
 \frac{k^2 b^2}{k_y^2 a^2}  
\nonumber\\
 &\ & 
-\frac{a^2b^2}{\lambda}V_\phi
\left(\frac{a^\prime}{a}-\frac{b^\prime}{b}\right)
+bb^\prime\frac{\phi^\prime}{\lambda^2}
\frac{b^2}{a^2}\frac{k^2}{k_y^2}
\frac{b^2f^2v^{\prime2}}{a^4}\frac{k_x^2}{k_y^2}
+2bb^\prime\frac{\phi^\prime}{\lambda^2}\frac{b^2}{a^2}\frac{k_x^2}{k_y^2}
\left(\frac{a^\prime}{a}-\frac{b^\prime}{b}\right)^2
 \nonumber\\
 &\ & 
\left.
-\frac{2}{\lambda}ff_\phi v^{\prime2}\frac{b^2}{a^2}\frac{b^\prime}{b}
+\frac{2}{\lambda}\frac{b^2}{a^2}ff_\phi v^{\prime2}
\left(\frac{k_x^2}{k^2}-\frac{1}{2}\right)
\left(\frac{a^\prime}{a}-\frac{b^\prime}{b}\right)
\right]
\left(G\delta\phi^*+G^*\delta\phi\right)
 \ .
\end{eqnarray}
Or, using variables $\bar{G},\bar{\delta\phi}$, we have
\begin{eqnarray}
\sqrt{2}L^{G\phi} &=& 
\left[
-\frac{b^2}{2a^4}f^2v^{\prime2}\frac{k_x^2}{k_y^2}\frac{\phi^\prime}{\lambda}
+2\frac{b^\prime}{b}\frac{b^2}{a^2}\frac{k_x^2}{k_y^2}
\frac{\phi^\prime}{\lambda}
\left(\frac{a^\prime}{a}-\frac{b^\prime}{b}\right)\right.
\nonumber\\
&\ & 
-\frac{\phi^\prime}{\lambda^2}\left(2a^2 V
+\frac{f^2v^{'2}}{2a^2}\frac{k_x^2}{k^2} 
+4\frac{k_x^2}{k_y^2}\frac{b^{\prime2}}{a^2}\right) 
\left(\frac{a^\prime}{a}-\frac{b^\prime}{b}\right)
 \frac{k^2 b^2}{k_y^2a^2}  
\nonumber\\
 &\ & 
-\frac{a^2}{\lambda}V_\phi
\left(\frac{a^\prime}{a}-\frac{b^\prime}{b}\right)
+\frac{b^\prime}{b}\frac{\phi^\prime}{\lambda^2}
\frac{b^2}{a^2}\frac{k^2}{k_y^2}
\frac{b^2f^2v^{\prime2}}{a^4}\frac{k_x^2}{k_y^2}
+2\frac{b^\prime}{b}\frac{\phi^\prime}{\lambda^2}\frac{b^2}{a^2}\frac{k_x^2}{k_y^2}
\left(\frac{a^\prime}{a}-\frac{b^\prime}{b}\right)^2
 \nonumber\\
 &\ & 
\left.
-\frac{2}{\lambda}
\frac{ff_\phi v^{\prime2}}{a^2}\frac{b^\prime}{b}
+\frac{2}{\lambda}\frac{ff_\phi v^{\prime2}}{a^2}
\left(\frac{k_x^2}{k^2}-\frac{1}{2}\right)
\left(\frac{a^\prime}{a}-\frac{b^\prime}{b}\right)
\right]
\left(\bar{G}\bar{\delta\phi}^*+\bar{G}^*\bar{\delta\phi}\right)
 \ .
\end{eqnarray}
The terms containing $\delta\phi $ and $J$ are given by
\begin{eqnarray}
L^{\phi J} &=& 
\frac{k_xk_y}{2k^2}f^2v^{\prime 2}
\left\{\frac{\phi\prime}{\lambda}
\frac{b^2}{a^2}\frac{k^2}{k_y^2}-2\frac{f_\phi}{f}\right\}
\left(J^\prime\delta\phi^*+{J^*}^\prime\delta\phi\right)
\nonumber\\
&\ & 
+\left(a^2b^2V+\frac{b^2}{2a^2}f^2v^{\prime2}\frac{k_x^2}{k^2}
+4\frac{b^2b^{\prime2}}{a^2}\frac{k_x^2}{k_y^2}\right)
\frac{v^\prime}{\lambda}\frac{f^2}{a^2}\frac{k_x}{k_y}
\frac{\phi^\prime}{\lambda}
\frac{k^2 b^2}{k_y^2a^2} 
\left(J\delta\phi^*+J^*\delta\phi\right)
 \nonumber\\
&\ & 
+\left\{\frac{1}{2}\left(\frac{b^2}{a^2}ff_\phi v^{\prime2}+a^2b^2V_\phi\right)
-\frac{b^2}{a^2}ff_\phi v^{\prime2}\frac{k_x^2}{k^2}
-2\frac{b^3b^\prime}{a^2}\frac{k_x^2}{k_y^2}\phi^\prime
\right\}\frac{v^\prime}{\lambda}\frac{f^2}{a^2}\frac{k_x}{k_y}
\left(J\delta\phi^*+J^*\delta\phi\right)
\nonumber\\
&\ & 
+\frac{b^2v^\prime}{2}\frac{\phi^\prime}{\lambda}\frac{f^2}{a^2}\frac{k_x}{k_y}
\left(J{\delta\phi^*}^\prime+J^*\delta\phi^\prime\right)
 \ . 
\end{eqnarray}
The above action can be rewritten by using variables 
$\bar{\delta\phi}=b\delta\phi$ and $\bar{J}=\frac{f|k_x|}{k}J$ as
\begin{eqnarray}
bL^{\phi J} &=& 
\frac{k_y}{2k}fv^{\prime 2}
\left\{\frac{\phi\prime}{\lambda}
\frac{b^2}{a^2}\frac{k^2}{k_y^2}-2\frac{f_\phi}{f}\right\}
\left(\bar{J}^\prime\bar{\delta\phi}^*+\bar{J^*}^\prime\bar{\delta\phi}\right)
\nonumber\\
&\ & 
\frac{k_y}{2k^2}f^2v^{\prime 2}\left(\frac{k}{f}\right)^\prime
\left\{\frac{\phi\prime}{\lambda}
\frac{b^2}{a^2}\frac{k^2}{k_y^2}-2\frac{f_\phi}{f}\right\}
\left(\bar{J}\bar{\delta\phi}^*+\bar{J}^*\bar{\delta\phi}\right)
\nonumber\\
&\ & 
+\left(a^2b^2V+\frac{b^2}{2a^2}f^2v^{\prime2}\frac{k_x^2}{k^2}
+4\frac{b^2b^{\prime2}}{a^2}\frac{k_x^2}{k_y^2}\right)
\frac{v^\prime}{\lambda}\frac{f}{a^2}\frac{k}{k_y}
\frac{\phi^\prime}{\lambda}
\frac{k^2 b^2}{k_y^2a^2} 
\left(\bar{J}\bar{\delta\phi}^*+\bar{J}^*\bar{\delta\phi}\right)
 \nonumber\\
&\ & 
+\left\{\frac{1}{2}\left(\frac{b^2}{a^2}ff_\phi v^{\prime2}+a^2b^2V_\phi\right)
-\frac{b^2}{a^2}ff_\phi v^{\prime2}\frac{k_x^2}{k^2}
-2\frac{b^3b^\prime}{a^2}\frac{k_x^2}{k_y^2}\phi^\prime
\right\}\frac{v^\prime}{\lambda}\frac{f}{a^2}\frac{k}{k_y}
\left(\bar{J}\bar{\delta\phi}^*+\bar{J}^*\bar{\delta\phi}\right)
\nonumber\\
&\ & 
+\frac{bb^\prime v^\prime}{2}\frac{\phi^\prime}{\lambda}\frac{f}{a^2}
\frac{k}{k_y}
\left(\bar{J}\bar{\delta\phi}^*+\bar{J}^*\bar{\delta\phi}\right)
+\frac{b^2v^\prime}{2}\frac{\phi^\prime}{\lambda}\frac{f}{a^2}\frac{k}{k_y}
\left(\bar{J}\bar{\delta\phi^*}^\prime+\bar{J}^*\bar{\delta\phi}^\prime\right)
 \ .
\end{eqnarray}
Finally, the terms containing $J$ and $G$ reads
\begin{eqnarray}
L^{JG}&=&
\left\{\frac{k_xk_y}{2k^2}f^2v^\prime
+\frac{k_xk_y}{2k^2}\frac{f^2v^\prime}{\lambda}
\left(\frac{a^\prime}{a}-\frac{b^\prime}{b}\right)
+bb^\prime\frac{v^\prime}{\lambda}\frac{f^2}{a^2}\frac{k_x}{k_y}
\right\}
\left(J^\prime G^*+{J^*}^\prime G\right)
\nonumber\\
&\ &
+\left(a^2b^2V+\frac{b^2}{2a^2}f^2v^{\prime2}\frac{k_x^2}{k^2}\right)
\frac{v^\prime}{\lambda}\frac{f^2}{a^2}\frac{k_x}{k_y}
\left(JG^*+J^*G\right)
+bb^\prime\frac{k_x}{k_y}
\left(\frac{v^\prime}{\lambda}\frac{f^2}{a^2}\right)^\prime
\left(JG^*+J^*G\right)
\nonumber\\
&\ &
+\left(a^2b^2V+\frac{b^2}{2a^2}f^2v^{\prime2}\frac{k_x^2}{k^2}
+4\frac{b^2b^{\prime2}}{a^2}\frac{k_x^2}{k_y^2}\right)
\frac{v^\prime}{\lambda}\frac{f^2}{a^2}\frac{k_x}{k_y}\frac{1}{\lambda}
\left(\frac{a^\prime}{a}-\frac{b^\prime}{b}\right)
\left(JG^*+J^*G\right)
 \ .
\end{eqnarray}
Using variables $\bar{G}=\sqrt{2}bG,\bar{J}=\frac{f|k_x|}{k}J$, we obtain
\begin{eqnarray}
\sqrt{2}bL^{JG} &=& 
\left\{\frac{k_y}{2k}fv^\prime
+\frac{k_y}{2k}\frac{fv^\prime}{\lambda}
\left(\frac{a^\prime}{a}-\frac{b^\prime}{b}\right)
+bb^\prime\frac{v^\prime}{\lambda}\frac{f}{a^2}\frac{k}{k_y}
\right\}
\left(\bar{J}^\prime \bar{G}^*+\bar{J^*}^\prime\bar{G}\right)
\nonumber\\
&\ &
+\left\{\frac{k_y}{2k^2}f^2v^\prime
+\frac{k_y}{2k^2}\frac{f^2v^\prime}{\lambda}
\left(\frac{a^\prime}{a}-\frac{b^\prime}{b}\right)
+bb^\prime\frac{v^\prime}{\lambda}\frac{f^2}{a^2}\frac{1}{k_y}
\right\}\left(\frac{k}{f}\right)^\prime 
\left(\bar{J} \bar{G}^*+\bar{J^*} \bar{G}\right)
\nonumber\\
&\ &
+\left(a^2b^2V+\frac{b^2}{2a^2}f^2v^{\prime2}\frac{k_x^2}{k^2}\right)
\frac{v^\prime}{\lambda}\frac{f}{a^2}\frac{k}{k_y}
\left(\bar{J}\bar{G}^*+\bar{J}^*\bar{G}\right)
+\frac{bb^\prime}{f}\frac{k}{k_y}
\left(\frac{v^\prime}{\lambda}\frac{f^2}{a^2}\right)^\prime
\left(\bar{J}\bar{G}^*+\bar{J}^*\bar{G}\right)
\nonumber\\
&\ &
+\left(a^2b^2V+\frac{b^2}{2a^2}f^2v^{\prime2}\frac{k_x^2}{k^2}
+4\frac{b^2b^{\prime2}}{a^2}\frac{k_x^2}{k_y^2}\right)
\frac{v^\prime}{\lambda}\frac{f}{a^2}\frac{k}{k_y}\frac{1}{\lambda}
\left(\frac{a^\prime}{a}-\frac{b^\prime}{b}\right)
\left(\bar{J}\bar{G}^*+\bar{J}^*\bar{G}\right)
 \ .
\end{eqnarray}
Thus, we have obtained the action for the physical variables $G$, $J$, and $\delta \phi$. 
It is useful to check if $J$, $G$ and $\delta \phi$ decouple to each other
in the special case $k_y=0$ for which there exists rotational symmetry in $y-z$ plane.
 Indeed, we find
\begin{eqnarray}
L^{G\phi} = L^{\phi J} = L^{GJ} = 0 \ .
\end{eqnarray}
The decoupling occurs because the helicity conserves in this special case.
The other parts of action become
\begin{eqnarray}
L^{GG} &=&  \frac{1}{2}|\bar{G}^{'}|^2 
+\frac{1}{2} \left[-k_x^2+\frac{b^{''}}{b} ~\right] |\bar{G}|^2 ,\\
L^{\phi\phi} &=& \frac{1}{2} |\bar{\delta\phi}^{\prime}|^2 
 +\frac{1}{2} \bigg[-k_x^2+\frac{b^{''}}{b}-a^2V_{\phi\phi}
-\frac{v^{\prime^2}}{a^2}\left(3f_{\phi}^2-ff_{\phi\phi}\right) 
-\frac{b^2}{b^{\prime2}}\phi ^{'2}
\left(a^2V+\frac{f^2v^{'2}}{2a^2} \right) \nonumber\\
 &\ & \hspace{6em} 
-2\frac{b}{b^\prime}\phi ^{'}
\left(a^2V_{\phi}-\frac{ff_{\phi}v^{'2}}{a^2} \right) 
 \bigg] |\bar{\delta\phi}|^2  \ ,\\
L^{JJ} &=& 
\frac{1}{2} |\bar{J}^{'}|^2 +\frac{1}{2}\left[ -k_x^2+\frac{f^{''}}{f} 
-2\frac{f^2v^{'2}}{a^2} \right] |\bar{J}|^2  \ .
\end{eqnarray}


\begin{thebibliography}{99}

%\cite{Komatsu:2010fb}
\bibitem{Komatsu:2010fb}
  E.~Komatsu {\it et al.},
  %``Seven-Year Wilkinson Microwave Anisotropy Probe (WMAP) Observations:
  %Cosmological Interpretation,''
  arXiv:1001.4538 [astro-ph.CO].
  %%CITATION = ARXIV:1001.4538;%%

%\cite{Maldacena:2002vr}
\bibitem{Maldacena:2002vr}
  J.~M.~Maldacena,
  %``Non-Gaussian features of primordial fluctuations in single field
  %inflationary models,''
  JHEP {\bf 0305}, 013 (2003)
  [arXiv:astro-ph/0210603].
  %%CITATION = JHEPA,0305,013;%%
  
%\cite{Wald:1983ky}
\bibitem{Wald:1983ky}
  R.~M.~Wald,
  %``Asymptotic behavior of homogeneous cosmological models in the presence of a
  %positive cosmological constant,''
  Phys.\ Rev.\  D {\bf 28}, 2118 (1983).
  %%CITATION = PHRVA,D28,2118;%%

%\cite{Eriksen:2003db}
\bibitem{Eriksen:2003db}
  H.~K.~Eriksen, F.~K.~Hansen, A.~J.~Banday, K.~M.~Gorski and P.~B.~Lilje,
  %``Asymmetries in the CMB anisotropy field,''
  Astrophys.\ J.\  {\bf 605}, 14 (2004)
  [Erratum-ibid.\  {\bf 609}, 1198 (2004)]
  [arXiv:astro-ph/0307507];
  %%CITATION = ASJOA,605,14;%%
  F.~K.~Hansen, A.~J.~Banday and K.~M.~Gorski,
  %``Testing the cosmological principle of isotropy: local power spectrum
  %estimates of the WMAP data,''
  Mon.\ Not.\ Roy.\ Astron.\ Soc.\  {\bf 354}, 641 (2004)
  [arXiv:astro-ph/0404206];
  %%CITATION = MNRAA,354,641;%%
  T.~R.~Jaffe, A.~J.~Banday, H.~K.~Eriksen, K.~M.~Gorski and F.~K.~Hansen,
  %``Evidence of vorticity and shear at large angular scales in the WMAP  data:
  %A violation of cosmological isotropy?,''
  Astrophys.\ J.\  {\bf 629}, L1 (2005)
  [arXiv:astro-ph/0503213];
  %%CITATION = ASJOA,629,L1;%%
  H.~K.~Eriksen, A.~J.~Banday, K.~M.~Gorski, F.~K.~Hansen and P.~B.~Lilje,
  %``Hemispherical power asymmetry in the three-year Wilkinson Microwave
  %Anisotropy Probe sky maps,''
  Astrophys.\ J.\  {\bf 660}, L81 (2007)
  [arXiv:astro-ph/0701089];
  %%CITATION = ASJOA,660,L81;%%
  A.~de Oliveira-Costa, M.~Tegmark, M.~Zaldarriaga and A.~Hamilton,
  %``The significance of the largest scale CMB fluctuations in WMAP,''
  Phys.\ Rev.\  D {\bf 69}, 063516 (2004)
  [arXiv:astro-ph/0307282];
  %%CITATION = PHRVA,D69,063516;%%
  K.~Land and J.~Magueijo,
  %``The axis of evil,''
  Phys.\ Rev.\ Lett.\  {\bf 95}, 071301 (2005)
  [arXiv:astro-ph/0502237];
  %%CITATION = PRLTA,95,071301;%%
  K.~Land and J.~Magueijo,
  %``The Axis of Evil revisited,''
  Mon.\ Not.\ Roy.\ Astron.\ Soc.\  {\bf 378}, 153 (2007)
  [arXiv:astro-ph/0611518];
  %%CITATION = MNRAA,378,153;%%
  C.~Copi, D.~Huterer, D.~Schwarz and G.~Starkman,
  %``The Uncorrelated Universe: Statistical Anisotropy and the Vanishing Angular
  %Correlation Function in WMAP Years 1-3,''
  Phys.\ Rev.\  D {\bf 75}, 023507 (2007)
  [arXiv:astro-ph/0605135];
  %%CITATION = PHRVA,D75,023507;%%
  J.~Hoftuft, H.~K.~Eriksen, A.~J.~Banday, K.~M.~Gorski, F.~K.~Hansen and P.~B.~Lilje,
  %``Increasing evidence for hemispherical power asymmetry in the five-year WMAP
  %data,''
  Astrophys.\ J.\  {\bf 699}, 985 (2009)
  [arXiv:0903.1229 [astro-ph.CO]];
  %%CITATION = ASJOA,699,985;%%
  P.~K.~Samal, R.~Saha, P.~Jain and J.~P.~Ralston,
  %``Signals of Statistical Anisotropy in WMAP Foreground-Cleaned Maps,''
  Mon.\ Not.\ Roy.\ Astron.\ Soc.\  {\bf 396}, 511 (2009)
  [arXiv:0811.1639 [astro-ph]].
  %%CITATION = MNRAA,396,511;%%

 %\cite{Gordon:2005ai}
\bibitem{Gordon:2005ai}
  C.~Gordon, W.~Hu, D.~Huterer and T.~M.~Crawford,
  %``Spontaneous Isotropy Breaking: A Mechanism for CMB Multipole Alignments,''
  Phys.\ Rev.\  D {\bf 72}, 103002 (2005)
  [arXiv:astro-ph/0509301];
  %%CITATION = PHRVA,D72,103002;%%
  C.~Armendariz-Picon,
  %``Creating Statistically Anisotropic and Inhomogeneous Perturbations,''
  JCAP {\bf 0709}, 014 (2007)
  [arXiv:0705.1167 [astro-ph]];
  %%CITATION = JCAPA,0709,014;%%
  D.~C.~Rodrigues,
  %``Anisotropic Cosmological Constant and the CMB Quadrupole Anomaly,''
  Phys.\ Rev.\  D {\bf 77}, 023534 (2008)
  [arXiv:0708.1168 [astro-ph]];
  %%CITATION = PHRVA,D77,023534;%%
  C.~Y.~Tseng and M.~B.~Wise,
  %``Inflaton Two-Point Correlation in the Presence of a Cosmic String,''
  Phys.\ Rev.\  D {\bf 80}, 103512 (2009)
  [arXiv:0908.0543 [astro-ph.CO]];
  %%CITATION = PHRVA,D80,103512;%%
  Y.~Shtanov and H.~Pyatkovska,
  %``Statistical anisotropy in the inflationary universe,''
  Phys.\ Rev.\  D {\bf 80}, 023521 (2009)
  [arXiv:0904.1887 [gr-qc]];
  %%CITATION = PHRVA,D80,023521;%%
  X.~Gao,
  %``Can Relic Superhorizon Inhomogeneities be Responsible for Large-Scale CMB
  %Anomalies?,''
  arXiv:0903.1412 [astro-ph.CO];
  %%CITATION = ARXIV:0903.1412;%%
  R.~Battye and A.~Moss,
  %``Anisotropic dark energy and CMB anomalies,''
  Phys.\ Rev.\  D {\bf 80}, 023531 (2009)
  [arXiv:0905.3403 [astro-ph.CO]].
  %%CITATION = PHRVA,D80,023531;%%
 %\cite{Yokoyama:2008xw}
\bibitem{Yokoyama:2008xw}
  S.~Yokoyama and J.~Soda,
  %``Primordial statistical anisotropy generated at the end of inflation,''
  JCAP {\bf 0808}, 005 (2008);
  %%CITATION = JCAPA,0808,005;%%
  
%\cite{Dimopoulos:2009vu}
\bibitem{Dimopoulos:2009vu}
 K.~Dimopoulos, D.~H.~Lyth and Y.~Rodriguez,
  %``Statistical anisotropy of the curvature perturbation from vector field
  %perturbations,''
  arXiv:0809.1055 [astro-ph];
  %%CITATION = ARXIV:0809.1055;%%
  T.~Kahniashvili, G.~Lavrelashvili and B.~Ratra,
  %``CMB Temperature Anisotropy from Broken Spatial Isotropy due to an
  %Homogeneous Cosmological Magnetic Field,''
  Phys.\ Rev.\  D {\bf 78}, 063012 (2008);
  %%CITATION = PHRVA,D78,063012;%% 
  K.~Dimopoulos, M.~Karciauskas and J.~M.~Wagstaff,
  %``Vector Curvaton without Instabilities,''
  Phys.\ Lett.\  B {\bf 683}, 298 (2010)
  [arXiv:0909.0475 [hep-ph]].
  %%CITATION = PHLTA,B683,298;%%
  
%\cite{Dimastrogiovanni:2010sm}
\bibitem{Dimastrogiovanni:2010sm}
  E.~Dimastrogiovanni, N.~Bartolo, S.~Matarrese and A.~Riotto,
  %``Non-Gaussianity and statistical anisotropy from vector field populated
  %inflationary models,''
  arXiv:1001.4049 [astro-ph.CO].
  %%CITATION = ARXIV:1001.4049;%%
    
%\cite{ValenzuelaToledo:2009af}
\bibitem{ValenzuelaToledo:2009af}
  C.~A.~Valenzuela-Toledo, Y.~Rodriguez and D.~H.~Lyth,
  %``Non-gaussianity at tree- and one-loop levels from vector field
  %perturbations,''
  Phys.\ Rev.\  D {\bf 80}, 103519 (2009)
  [arXiv:0909.4064 [astro-ph.CO]].
  %%CITATION = PHRVA,D80,103519;%%
  C.~A.~Valenzuela-Toledo and Y.~Rodriguez,
  %``Non-gaussianity from the trispectrum and vector field perturbations,''
  Phys.\ Lett.\  B {\bf 685}, 120 (2010)
  [arXiv:0910.4208 [astro-ph.CO]].
  %%CITATION = PHLTA,B685,120;%%
        
%\cite{Ford:1989me}
\bibitem{Ford:1989me}
  L.~H.~Ford,
  %``INFLATION DRIVEN BY A VECTOR FIELD,''
  Phys.\ Rev.\  D {\bf 40}, 967 (1989).
  %%CITATION = PHRVA,D40,967;%%

%\cite{Kaloper:1991rw}
\bibitem{Kaloper:1991rw}
  N.~Kaloper,
  %``Lorentz Chern-Simons terms in Bianchi cosmologies and the cosmic no hair
  %conjecture,''
  Phys.\ Rev.\  D {\bf 44}, 2380 (1991).
  %%CITATION = PHRVA,D44,2380;%%  
  
%\cite{Kawai:1998bn}
\bibitem{Kawai:1998bn}
  S.~Kawai and J.~Soda,
  %``Non-singular Bianchi type I cosmological solutions from 1-loop  superstring
  %effective action,''
  Phys.\ Rev.\  D {\bf 59}, 063506 (1999)
  [arXiv:gr-qc/9807060].
  %%CITATION = PHRVA,D59,063506;%%
  
%\cite{Barrow:2005qv}
\bibitem{Barrow:2005qv}
  J.~D.~Barrow and S.~Hervik,
  %``Anisotropically inflating universes,''
  Phys.\ Rev.\  D {\bf 73}, 023007 (2006)
  [arXiv:gr-qc/0511127].
  %%CITATION = PHRVA,D73,023007;%%     

%\cite{Barrow:2009gx}
\bibitem{Barrow:2009gx}
  J.~D.~Barrow and S.~Hervik,
  %``Simple Types of Anisotropic Inflation,''
  Phys.\ Rev.\  D {\bf 81}, 023513 (2010)
  [arXiv:0911.3805 [gr-qc]].
  %%CITATION = PHRVA,D81,023513;%%
  
%\cite{Campanelli:2009tk}
\bibitem{Campanelli:2009tk}
  L.~Campanelli,
  %``A Model of Universe Anisotropization,''
  Phys.\ Rev.\  D {\bf 80}, 063006 (2009)
  [arXiv:0907.3703 [astro-ph.CO]].
  %%CITATION = PHRVA,D80,063006;%%
  
%\cite{Golovnev:2008cf}
\bibitem{Golovnev:2008cf}
  A.~Golovnev, V.~Mukhanov and V.~Vanchurin,
  %``Vector Inflation,''
  JCAP {\bf 0806}, 009 (2008)
  [arXiv:0802.2068 [astro-ph]];
  %%CITATION = JCAPA,0806,009;%%
  
  %\cite{Kanno:2008gn}
\bibitem{Kanno:2008gn}
  S.~Kanno, M.~Kimura, J.~Soda and S.~Yokoyama,
  %``Anisotropic Inflation from Vector Impurity,''
  JCAP {\bf 0808}, 034 (2008).
  %%CITATION = JCAPA,0808,034;%%

 %\cite{Ackerman:2007nb}
\bibitem{Ackerman:2007nb}
  L.~Ackerman, S.~M.~Carroll and M.~B.~Wise,
  %``Imprints of a Primordial Preferred Direction on the Microwave Background,''
  Phys.\ Rev.\  D {\bf 75}, 083502 (2007).
  %%CITATION = PHRVA,D75,083502;%%
   
%\cite{Himmetoglu:2008zp}
\bibitem{Himmetoglu:2008zp}
  B.~Himmetoglu, C.~R.~Contaldi and M.~Peloso,
  %``Instability of anisotropic cosmological solutions supported by vector
  %fields,''
  arXiv:0809.2779 [astro-ph];
  %%CITATION = ARXIV:0809.2779;%%
  B.~Himmetoglu, C.~R.~Contaldi and M.~Peloso,
  %``Instability of the ACW model, and problems with massive vectors during
  %inflation,''
  arXiv:0812.1231 [astro-ph];
  %%CITATION = ARXIV:0812.1231;%%
  B.~Himmetoglu, C.~R.~Contaldi and M.~Peloso,
  %``Ghost instabilities of cosmological models with vector fields nonminimally
  %coupled to the curvature,''
  Phys.\ Rev.\  D {\bf 80}, 123530 (2009)
  [arXiv:0909.3524 [astro-ph.CO]].
  %%CITATION = PHRVA,D80,123530;%%

%\cite{Watanabe:2009ct}
\bibitem{Watanabe:2009ct}
  M.~a.~Watanabe, S.~Kanno and J.~Soda,
  %``Inflationary Universe with Anisotropic Hair,''
  Phys.\ Rev.\ Lett.\  {\bf 102}, 191302 (2009)
  [arXiv:0902.2833 [hep-th]].
  %%CITATION = PRLTA,102,191302;%%    

%\cite{Kanno:2009ei}
\bibitem{Kanno:2009ei}
  S.~Kanno, J.~Soda and M.~a.~Watanabe,
  %``Cosmological Magnetic Fields from Inflation and Backreaction,''
  JCAP {\bf 0912}, 009 (2009)
  [arXiv:0908.3509 [astro-ph.CO]].
  %%CITATION = JCAPA,0912,009;%%
 
%\cite{Pullen:2007tu}
\bibitem{Pullen:2007tu}
  A.~R.~Pullen and M.~Kamionkowski,
  %``Cosmic Microwave Background Statistics for a Direction-Dependent Primordial
  %Power Spectrum,''
  Phys.\ Rev.\  D {\bf 76}, 103529 (2007);
  %%CITATION = PHRVA,D76,103529;%%  
  N.~E.~Groeneboom and H.~K.~Eriksen,
  %``Bayesian analysis of sparse anisotropic universe models and application to
  %the 5-yr WMAP data,''
  Astrophys.\ J.\  {\bf 690}, 1807 (2009);
  %%CITATION = ASJOA,690,1807;%%
  C.~Armendariz-Picon and L.~Pekowsky,
  %``Bayesian Limits on Primordial Isotropy Breaking,''
  arXiv:0807.2687 [astro-ph].
  %%CITATION = ARXIV:0807.2687;%%
               
%\cite{Baumann:2008aq}
\bibitem{Baumann:2008aq}
  D.~Baumann {\it et al.}  [CMBPol Study Team Collaboration],
  %``CMBPol Mission Concept Study: Probing Inflation with CMB Polarization,''
  arXiv:0811.3919 [astro-ph];
  %%CITATION = ARXIV:0811.3919;%%
    
%\cite{Gluscevic:2010vv}
\bibitem{Gluscevic:2010vv}
  V.~Gluscevic and M.~Kamionkowski,
  %``Testing Parity-Violating Mechanisms with Cosmic Microwave Background
  %Experiments,''
  arXiv:1002.1308 [astro-ph.CO].
  %%CITATION = ARXIV:1002.1308;%%
  
%\cite{Seto:2001qf}
\bibitem{Seto:2001qf}
  N.~Seto, S.~Kawamura and T.~Nakamura,
  %``Possibility of direct measurement of the acceleration of the universe
  %using 0.1-Hz band laser interferometer gravitational wave antenna in
  %space,''
  Phys.\ Rev.\ Lett.\  {\bf 87}, 221103 (2001)
  [arXiv:astro-ph/0108011].
  %%CITATION = PRLTA,87,221103;%%
    
%\cite{Tomita:1985me}
\bibitem{Tomita:1985me}
  K.~Tomita and M.~Den,
  %``Gauge Invariant Perturbations In Anisotropic Homogeneous Cosmological
  %Models,''
  Phys.\ Rev.\  D {\bf 34}, 3570 (1986).
  %%CITATION = PHRVA,D34,3570;%% 
  
%\cite{Dunsby:1993fg}
\bibitem{Dunsby:1993fg}
  P.~K.~S.~Dunsby,
  %``Gauge Invariant Perturbations Of Anisotropic Cosmological Models,''
  Phys.\ Rev.\  D {\bf 48}, 3562 (1993).
  %%CITATION = PHRVA,D48,3562;%%
  
%\cite{Noh:1987vk}
\bibitem{Noh:1987vk}
  H.~Noh and J.~C.~Hwang,
  %``Perturbations of an anisotropic space-time: Formulation,''
  Phys.\ Rev.\  D {\bf 52}, 1970 (1995).
  %%CITATION = PHRVA,D52,1970;%%
      
%\cite{Pereira:2007yy}
\bibitem{Pereira:2007yy}
  T.~S.~Pereira, C.~Pitrou and J.~P.~Uzan,
  %``Theory of cosmological perturbations in an anisotropic universe,''
  JCAP {\bf 0709}, 006 (2007)
  [arXiv:0707.0736 [astro-ph]];
  %%CITATION = JCAPA,0709,006;%% 
  C.~Pitrou, T.~S.~Pereira and J.~P.~Uzan,
  %``Predictions from an anisotropic inflationary era,''
  JCAP {\bf 0804}, 004 (2008)
  [arXiv:0801.3596 [astro-ph]].
  %%CITATION = JCAPA,0804,004;%%
  
%\cite{Gumrukcuoglu:2007bx}
\bibitem{Gumrukcuoglu:2007bx}
  A.~E.~Gumrukcuoglu, C.~R.~Contaldi and M.~Peloso,
  %``Inflationary perturbations in anisotropic backgrounds and their imprint on
  %the CMB,''
  JCAP {\bf 0711}, 005 (2007)
  [arXiv:0707.4179 [astro-ph]].
  %%CITATION = JCAPA,0711,005;%%
  
%\cite{Himmetoglu:2009mk}
\bibitem{Himmetoglu:2009mk}
  B.~Himmetoglu,
  %``Spectrum of Perturbations in Anisotropic Inflationary Universe with Vector
  %Hair,''
  arXiv:0910.3235 [astro-ph.CO].
  %%CITATION = ARXIV:0910.3235;%%  

%\cite{Dulaney:2010sq}
\bibitem{Dulaney:2010sq}
  T.~R.~Dulaney and M.~I.~Gresham,
  %``Primordial Power Spectra from Anisotropic Inflation,''
  arXiv:1001.2301 [astro-ph.CO].
  %%CITATION = ARXIV:1001.2301;%%
  
%\cite{Gumrukcuoglu:2010yc}
\bibitem{Gumrukcuoglu:2010yc}
  A.~E.~Gumrukcuoglu, B.~Himmetoglu and M.~Peloso,
  %``Scalar-Scalar, Scalar-Tensor, and Tensor-Tensor Correlators from
  %Anisotropic Inflation,''
  arXiv:1001.4088 [astro-ph.CO].
  %%CITATION = ARXIV:1001.4088;%%

%\cite{Martin:2007ue}
\bibitem{Martin:2007ue}
  J.~Martin and J.~Yokoyama,
  %``Generation of Large-Scale Magnetic Fields in Single-Field Inflation,''
  JCAP {\bf 0801}, 025 (2008).
  %%CITATION = JCAPA,0801,025;%%  

%\cite{Lue:1998mq}
\bibitem{Lue:1998mq}
  A.~Lue, L.~M.~Wang and M.~Kamionkowski,
  %``Cosmological signature of new parity-violating interactions,''
  Phys.\ Rev.\ Lett.\  {\bf 83}, 1506 (1999)
  [arXiv:astro-ph/9812088];
  %%CITATION = PRLTA,83,1506;%%
  S.~Alexander and J.~Martin,
  %``Birefringent gravitational waves and the consistency check of  inflation,''
  Phys.\ Rev.\  D {\bf 71}, 063526 (2005)
  [arXiv:hep-th/0410230];
  %%CITATION = PHRVA,D71,063526;%%
  M.~Satoh, S.~Kanno and J.~Soda,
  %``Circular Polarization of Primordial Gravitational Waves in String-inspired
  %Inflationary Cosmology,''
  Phys.\ Rev.\  D {\bf 77}, 023526 (2008);
  %%CITATION = PHRVA,D77,023526;%%
 M.~Satoh and J.~Soda,
  %``Higher Curvature Corrections to Primordial Fluctuations in Slow-roll
  %Inflation,''
  JCAP {\bf 0809}, 019 (2008);
  %%CITATION = JCAPA,0809,019;%%
  Y.~f.~Cai and Y.~S.~Piao,
  %``Probing Noncommutativity with Inflationary Gravitational Waves,''
  Phys.\ Lett.\  B {\bf 657}, 1 (2007)
  [arXiv:gr-qc/0701114];
  %%CITATION = PHLTA,B657,1;%%   
  C.~R.~Contaldi, J.~Magueijo and L.~Smolin,
  %``Anomalous CMB polarization and gravitational chirality,''
  Phys.\ Rev.\ Lett.\  {\bf 101}, 141101 (2008)
  [arXiv:0806.3082 [astro-ph]];
  %%CITATION = PRLTA,101,141101;%%
  T.~Takahashi and J.~Soda,
  %``Chiral Primordial Gravitational Waves from a Lifshitz Point,''
  Phys.\ Rev.\ Lett.\  {\bf 102}, 231301 (2009)
  [arXiv:0904.0554 [hep-th]].
  %%CITATION = PRLTA,102,231301;%%
                 
%\cite{Dechant:2008pb}
\bibitem{Dechant:2008pb}
  P.~P.~Dechant, A.~N.~Lasenby and M.~P.~Hobson,
  %``An anisotropic, non-singular early universe model leading to a realistic
  %cosmology,''
  Phys.\ Rev.\  D {\bf 79}, 043524 (2009)
  [arXiv:0809.4335 [gr-qc]].
  %%CITATION = PHRVA,D79,043524;%%
        
%\bibliography{apssamp}% Produces the bibliography via BibTeX.

\end{thebibliography}
\end{document}